\documentclass[11pt]{article}       
\usepackage{latex8}
\usepackage{times}
\usepackage{epsfig}
\usepackage{amssymb}
\usepackage{amsmath}
\usepackage{subfig}
\usepackage{wrapfig,lipsum,booktabs}
\usepackage{url}
\usepackage{color}
\usepackage{epstopdf}

\newcommand{\sqm}{\begin{pmatrix}}

\newcommand{\x}{\mathbf{x}}

\newcommand{\X}{\mathbf{X}}

\begin{document}
\title{On The Discovery of Social Roles in Large Scale Social Systems}
\author{Derek Doran \\
Department of Computer Science \& Engineering\\
Kno.e.sis Research Center \\
Wright State University, Dayton, OH \\
derek.doran@wright.edu
}
\maketitle

\begin{abstract}
The {\em social role} of a participant in a social system is a label conceptualizing 
the circumstances under which she interacts within it. They may be used as a theoretical
tool that explains why and how users participate in an online social system. 
Social role analysis also serves practical purposes, such as reducing the 
structure of complex systems to relationships among roles rather than alters, 
and enabling a comparison of social systems that emerge 
in similar contexts. This article presents a data-driven
approach for the discovery of social roles in large scale social systems. 
Motivated by an analysis of the present art, the method discovers roles 
by the conditional triad censuses of user ego-networks, which is a promising tool 
because they capture the degree to which basic social forces 
push upon a user to interact with others. Clusters of censuses, inferred 
from samples of large scale network carefully chosen to preserve local structural 
properties, define the social roles. The promise of the 
method is demonstrated by discussing and discovering the roles that emerge in
both Facebook and Wikipedia. The article
concludes with a discussion of the challenges and future opportunities in the 
discovery of social roles in large social systems. 
\end{abstract}

\section{Introduction and Motivation}
\label{sec:intro}
Why do people choose to participate and interact with others in a social system? 
This basic question lies at the heart of many sociological studies 
that examine the nature of interactions in a community. The question
is theoretically associated with the {\em social roles} of community
members, which is  
defined as a qualitative description capturing the circumstances
and reasons under which they choose to interact with others. 
The concept of a social role is fundamentally based on the notion of 
a user's {\em position} within a social network~\cite{gleave09,lorrain71}.
For example, with whom and how one decides to connect to others 
in a community is
associated with how they are perceived by others~\cite{brass98}, the power they hold~\cite{cook78}, and their ability to spread information 
and influence others~\cite{zaheer05}. As a concrete illustration, consider
a network of interactions among workers in a corporate office. Some 
workers have the social role ``manager" as defined by who they are connected
to socially: ``managers" are responsible for the work of the ``team members" he leads
and report to an ``executive". Any person in the corporate office network 
in a similar position, even if they report
to a different executive and managers a different team, is still perceived 
to have the social role ``manager". 

Extracting and understanding the social roles of a social system carries theoretical and practical importance. Theoretically, 
an analyst may integrate the social roles discovered in a social setting and
the context of these interactions to formulate a thesis about the reasons why and how people interact within the system. For example, 
consider the typical interactions that may occur within a generic corporate office as well as the connotations of being labeled a ``manager". 
Analysts could infer that ``managers" interact
with ``team members" based on the initiatives and projects assigned to them
by ``executives". They may be required to balance the demands placed on them by executives along with the needs of the team, and serve as a broker that 
filters information from corporate leaders to others in the organization. 
Practically, the delineation of users by their social role facilitates the  interpretation of complex social systems by simplifying their structure from
connections among users to between roles~\cite{borgatti89,scott11,white76}. 
It also enables meaningful studies of communities across time and 
context (e.g., different types corporate offices) by comparing the structure of interactions between roles that are common among them. 
For example, meta-analysis of the
social roles roles and the interactions among them roles across 
different groups can help designers create effective physical 
and digital spaces for communities and organizations to 
grow within~\cite{hautz10}.  Social role analysis is also  
useful to identify the types of users that may become
influential~\cite{gliwa13}, and even reveal latent social structures 
within the systems~\cite{laniado11}. 

This article presents a new method to discover the social roles that exist
in large scale online social systems. The methodology is motivated by an
analysis of the present art, which either:~(i) requires an analyst to presume the 
existence of roles beforehand; and/or~(ii) mines the roles using features
about the users and the structure of the system that may not have a basis
in social theory. The approach discovers social roles by clustering users 
by their {\em conditional triad census}, which is a 
vector capturing the types and orientations of three way 
relationships their ego-network is composed of. 
The method is applied to a network of interactions from an online social 
network (Facebook) and a collaborative editing platform (Wikipedia). An
analysis of the quality of the resulting clusters and the ego-network structure
of prototypical users demonstrate the utility of the proposed
method. The article concludes with a discussion about the many opportunities 
and challenges for future research in social role discovery for large scale
social systems. 

This article is organized as follows: 
Section~\ref{sec:rr} reviews and assesses existing methods for 
social role discovery in large scale social systems. 
Section~\ref{sec:mining}  introduces the concept
of a conditional triad census and the proposed methodology. 
Section~\ref{sec:analysis} analyzes the structure of the social 
roles mined from two large scale online social systems. 
Important challenges and opportunities that remain in the analysis
of social roles in large scale systems are presented in Section~\ref{sec:opp}.
Concluding remarks are offered in Section~\ref{sec:conc}.

\section{Discovering Social Roles}
\label{sec:rr}
Present methods to discover social roles in social systems 
may be classified into three types:~(i)
methods that define roles by notions of equivalence;~(ii)
methods that require the assertion of the roles existing in the
system prior to analysis; and~(iii) methods that define roles based
on patterns among user attributes and system interactions. 
This section provides an overview of each type
and their applicability to discover 
social roles in large scale systems.

\subsection{Equivalence based role discovery}
Longstanding methods to identify social roles are based on
finding users who are in ``equivalent" 
positions~\cite{white76,borgatti92notions,borgatti92regular,borgatti93}, 
which may be defined in one of three ways. Given 
an undirected network $G = (V,E)$ of users $V$ connected by a set of relations $E$, {\em structural
equivalence} requires two users $i$ and $j$ to be connected
to be exactly the same set of others. In other words, for
every relationship $(i,x) \in E$ that exists, the relation $(j,x)$
must also exist. Under this definition, a user's 
social role is precisely defined by the 
people that she is connected to. This strict definition
may not be useful in many settings because it is impossible for two users
whose distance is greater than two in a network
to fall under the same role. For example, two ``managers" in an office
that report to a common ``executive'' but have difference sets of 
subordinates are not structurally equivalent 
and would therefore not be classified under the same role. 

{\em Isomorphic equivalence} offers a broader definition
of equivalent network positions. An isomorphism among
two users in a network exist if there is a 
mapping $\pi: E(a) \to E(b)$ where $E(a)$ is the 
set of relationships held by user $a$ such that 
for every pair of users $a, b \in E$, we have 
$(a,b) \in E(a)$ if and only if 
$(\pi(a),\pi(b)) \in E(b)$. In other words, users $a$ 
and $b$ must have isomorphic {\em ego-networks}, which is a tuple
$(V_e,E_e)$ where $V_e$ is the set of 
all users in the $2^{nd}$ degree neighborhood of a user and $E_e$ represents
the directed relationships that bind the users in $V_e$ together. 
This suggests
that one could simply switch the location of user
$a$ and $b$ and their connectivity to others without disturbing 
the overall structure of the network. Practically, 
two ``managers" in a network that report to an ``executive" and
lead the same number of ``team members" would 
be isomorphically equivalent if the connectivity among  the
``team members" of the two ``managers" were isomorphic.
This equivalence definition thus captures a more
intuitive notion for ascribing a user's role in a social
system. A still broader class is {\em regular equivalence}, 
which requires the role of the alters of two users to be 
identical. Specifically,  
if $\mathcal{R}(x)$ is a function that assigns a user $x$
to a role, we say users $a$ and $b$ are regularly equivalent
if $\mathcal{R}(a) = \mathcal{R}(b)$
and if every user $n$ in the ego-network $N(a)$
of $a$  can be mapped to a user $m$ in the ego-network of $b$ 
such that $\mathcal{R}(n) = \mathcal{R}(m)$. For example, ``managers''
would be regularly equivalent so long as they both connect to 
``executives'' and ``team members''. Isomorphic
and regular equivalences may be identified by performing a blockmodeling
over the adjacency matrix of a social system~\cite{wasserman94}.

Notions of structural, isomorphic, and regular equivalence
are decades old theories that have been instrumental in 
many social network analyses~\cite{smith92,dimaggio86,cook83,wellman97,white97,erickson88}. 
More recent work have used these notions to study
international relationships across institutions~\cite{morselli12}, firms~\cite{pallotti11}, governments~\cite{zhou12,kick11}, 
and to study peer influences~\cite{fujimoto12}.
Isomorphic equivalence has been applied to hospitals within
referral networks~\cite{jung10} to discover closed 
communities of health services and hospitals
that carry identical areas of expertise. They are also
employed in the study of citation networks~\cite{tselykh13} to identify 
researchers within an organization that perform similar research and
offer similar domain expertise. Regular equivalences have been studied in 
networks of relations among gang members in urban settings~\cite{radil10} 
and of relations among cities across the world~\cite{alderson04}.
 
\subsection{Implied role discovery}

\begin{figure}
\centering
\includegraphics[width=120mm]{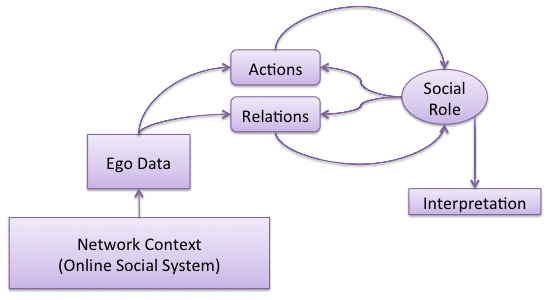}
\caption{Workflow for implied-role analysis}
\label{fig:pre}
\end{figure}

In implied role analysis, a researcher defines the
set of social roles users of a social system
are expected to exhibit before any data or structural
analysis commences. It is a qualitative, 
iterative process that generally follows the workflow of Figure~\ref{fig:pre}. 
Based on at-hand information about a social system, roles are first defined
based on the subset of functionality
allowed by the system that the user may  perform. 
For example, consider an online forum where users may decide
to browse conversations but never post, or can become
an administrator that edits and controls the behavior of 
others in the system. An analyst may therefore first define the social
roles {\em lurker} (one who never posts), {\em moderator}
(one who controls behavior), and {\em poster} (one
who contributes to conversations). With these roles assumed to 
exist, the analyst studies the actions of users and 
their relations with others. The initial definitions 
of the social roles are then iteratively
refined as evidence from the social system is collected. 

Implied role analysis is useful when a social
system is well understood, highly structured, and 
if the analyst wishes to understand the interactions
among users on the basis of the kinds of operations
they perform. For example, 
Nolker {\em et al.} tapped into their experiences with online
bulletin board systems to predefine members of a Usenet group into 
the roles {\em leader, motivator}, and {\em chatter}~\cite{nolker05}. 
They identified the behavioral attributes that are indicative of each role,
and labeled users exhibiting such behaviors in a log of the group's activity.
Golder {\em et al.} also studied Usenet groups but proposed a different
taxonomy of roles that include {\em celebrities, ranters, lurkers,
trolls,} and {\em newbies}~\cite{golder04}. They sifted through 
conversations across different Usenet groups to study
behaviors associated with each role. Gliwa {\em et al.} examined 
collections of online bloggers and defined the roles 
{\em selfish influential user, social influential user, selfish
influential blogger, social influential blogger, influential commentator,
standard commentator, not active}, and {\em standard blogger}~\cite{gliwa13}. 
Welser {\em et al.} defined four roles for Wikipedia users, namely
{\em substantive experts, technical editors, counter vandalism,} and 
{\em social networkers}~\cite{welser11}. They subsequently searched for
patterns about how users contribute and interact with others in order to 
classify the users falling in each role. 

\subsection{Data-driven role discovery}
\begin{figure}
\centering
\includegraphics[width=120mm]{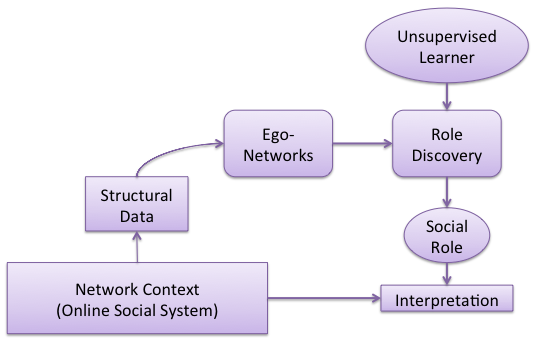}
\caption{Workflow for data-driven role analysis}
\label{fig:methme}
\end{figure}

A third type of approach is to infer social roles by the 
features of a dataset without pre-defining the roles that exist. 
These data-driven approaches, whose
workflow is summarized in Figure~\ref{fig:methme}, generally
considers features about 
users and the structure of their ego-networks in an 
unsupervised machine learning algorithm. Social roles are 
defined as the groups the algorithm places users into based on the
similarity of these features. 
Studies that apply unsupervised learners for social role
discovery vary in sophistication. For example, 
Hautz {\em et al.} categorized users in an online
community of jewelry designers by 
mapping whether their out- and in-degree
distributions and frequency of interactions to ``low" or
``high" levels~\cite{hautz10}. 
Zhu {\em et al.} use $k$-means clustering to identify user 
roles in a network of phone calls 
based on similar calling behaviors, ego-network 
clustering coefficients, and mean geodesic distances 
between users~\cite{zhu11}. Chan {\em et al.} discover roles by agglomerative hierarchical clustering with over fifty behavioral and structural 
features of users' across the post/reply network of many online forums~\cite{chan10}. White {\em et al.} use a mixed membership probabilistic model 
to identify roles across online forums using behavioral features and found
a number of possible assignments of users into groups~\cite{white12}. 
Rowe {\em et al.} use behavioral ontologies and semantic rules to 
automatically group online forum users into roles based on the content of their posts~\cite{rowe13}. Although data-driven approaches 
define similarity based on the structural features of ego-networks, this class of methods is not an approximation of equivalence based role discovery. This is because  data-driven methods may search for the similarity of two users based on many feature types that are not structural, including their personal attributes, 
their behaviors on the social system, and the content of their interactions with others. 

\subsection{Comparative analysis}
\label{sec:dda}


The recent availability of data about very large scale social systems,
typically collected from online social networks
(Facebook; Google+), social media (Twitter; Tumblr), 
and innovative information exchanges (Wikipedia; StackExchange)
enables the study of the social roles of users in systems that
have a world-wide reach. The massive scale of these systems necessitates
the need to evaluate current approaches for discovering social
roles, so that the most effective type given their size can be
identified. 

Equivalence based role discovery comprises a number of
well-studied, longstanding methods that has deep roots in 
sociological theory. Unfortunately, it may be infeasible
to precisely identify users falling into isomorphic or regular equivalence 
classes within large scale social systems.
This is because the problem of finding
isomorphic ego-networks is closely aligned to searching
for all motifs of arbitrary size within 
the network, and the problem of identifying 
regularly equivalent positions is related to searching for a 
$k$-coloring of $G$, with $k$ unknown a priori (both are 
NP-hard problems~\cite{khot01}).
Researchers still interested in identifying these 
equivalences in large systems must resort to 
numerical approximations based on quantitative notions
of structural similarity between two 
users that may be difficult to apply and analyze
in practice~\cite{newman,fan13many,jin11}. Thus, despite
the rich theory they are grounded within,  
technical challenges bar its adequate adaptation for large scale social systems. 

Implied role analyses carry fewer technical
challenges. This is because the most difficult aspect -
identifying the roles that exist - are predefined by an
analyst before trends in the data are considered. 
However, implied role analyses runs the risk
of using noisy signals in the data that appear by chance
as evidence for the roles they have predefined. 
Furthermore, it is possible for
separate analysts to define completely different sets of social
roles for the same system, which may confuse or conflict each other. 
For example, Nolker {\em et al.} places Usenet members into 
{\em leader, motivator}, and 
{\em chatter} roles~\cite{nolker05}. 
Are these roles compatible with the alternative
set of {\em celebrities, ranters, lurkers, trolls,} and {\em newbie} roles  proposed by Golder {\em et al.} for the same system~\cite{golder04}? 
It is unclear if one set of roles is more suitable than the other, 
or if the cross-product of the two
types of roles (e.g. {\em leader-celeberty} or {\em chatter-lurker}) 
is also a valid set of roles. Furthermore, the implied roles
tend to speak to the functionality or actions
that users of the social system undertake instead of reflecting 
the reasons why they participate in the system and the way 
they are structurally embedded within it. Thus, although
there are fewer technical challenges to run implied role analysis
over large scale social systems, the resulting roles may have
a weak relationship to sociological theory. 

Data-driven social role analysis may be 
a promising type of approach for the discovery of social roles in 
large-scale social systems. This is because modern day ``big data" technologies enable the collection of incredible amounts of information about each user, their connections with others in the social system, and the details or the content of their interactions.
Instead of assuming that specific kinds of social roles in the system
must exist, data-driven analyses apply data mining 
algorithms or learn data models from which the social roles of the system 
emerge. Such approaches let the data inform the analyst what 
social roles exist, rather than require a definition of the roles before 
studying the data. Fortunately, recent big data systems and
methods research enable the rapid mining and building of data models
from large social systems. For example, Zhang {\em et al.} 
tackle computations over real-world and virtual social interaction data by performing Tucker decompositions of a tensor representation of the interactions~\cite{zhang14}. A distributed learning algorithm
based on the MapReduce proposed by Tang {\em et al.} efficiently identifies the 
influencers and experts latent within large social systems~\cite{tang09}. 
Cambria {\em et al.} use a comparative analysis of the performance of 
multiple natural language processing algorithms to find patterns in the 
content of social interactions~\cite{cambria13}.  Giannakis {\em et al.} 
present a series of articles that describe how sensor signal processing 
algorithms may be adapted to operate over big and social data sets~\cite{giannakis14}. Malcom {\em et al.} even
developed a uniform programming interface so that non-experts can utilize
state-of-the-art big data technologies~\cite{malcom14} for social role
analysis. However, the relationship of such analyses with 
longstanding social theory varies considerably. This is because while some 
data mining algorithms and models encode aspects of social theory in their 
technical development, others were given no consideration to these theories in 
their development or make assumptions that are incompatible with past social 
science research for sake of model tractability. Furthermore, algorithms and models for social role analysis may use features that do not reflect aspects 
of social forces that drive users to embed themselves in the network in a 
specific way~\cite{chan10}.

\section{Triad-based Social Role Extraction}
\label{sec:mining}
In this section, a new data-driven approach
for extracting social roles from large social systems
is introduced\footnote{Parts of this method were presented 
at the First Workshop on Interaction and Exchange in Social Media
at the 2014 International Conference on Social 
Informatics~\cite{dyad14}.}. Based on the discussion in the previous section, it 
only considers features that have a grounding in social theory, namely the 
{\em conditional triads} that compose each user's ego-network. After network sampling and dimensionality reduction, 
$k$-means clustering is applied to the vectors to identify
social roles. Ego-networks falling closest to the centroid 
of each cluster is interpreted for role analysis. This section
describes what conditional
triads are, the triad-based representation of an ego-network,
the social systems used to illustrate the methodology, 
and the role extraction process.

\subsection{Conditional Triad Census}
\label{sec:census}
In social network analysis, a {\em triad} is a  
group of three individuals and the pairwise interactions
among them~\cite{simmel50}. 
They are the smallest sociological unit from
which the dynamics of a multi-person relationship can be observed, and hence, 
are considered to be the atomic unit of a social 
network~\cite{faust08,wasserman72,davis72}. For example, third actors may act as a
moderating force that can resolve conflicts among 
two others~\cite{brass98}. They may also sabotage an existing relationship or 
induce a feeling of unwelcomeness to a specific alter~\cite{baum06}. 
Such observations have been used to develop theories that associate the 
configuration of a triad to specific underlying effects that promote
specific kinds of social interactions~\cite{holland78,batjargal07}. 

\begin{figure}
\centering
\includegraphics[width=150mm]{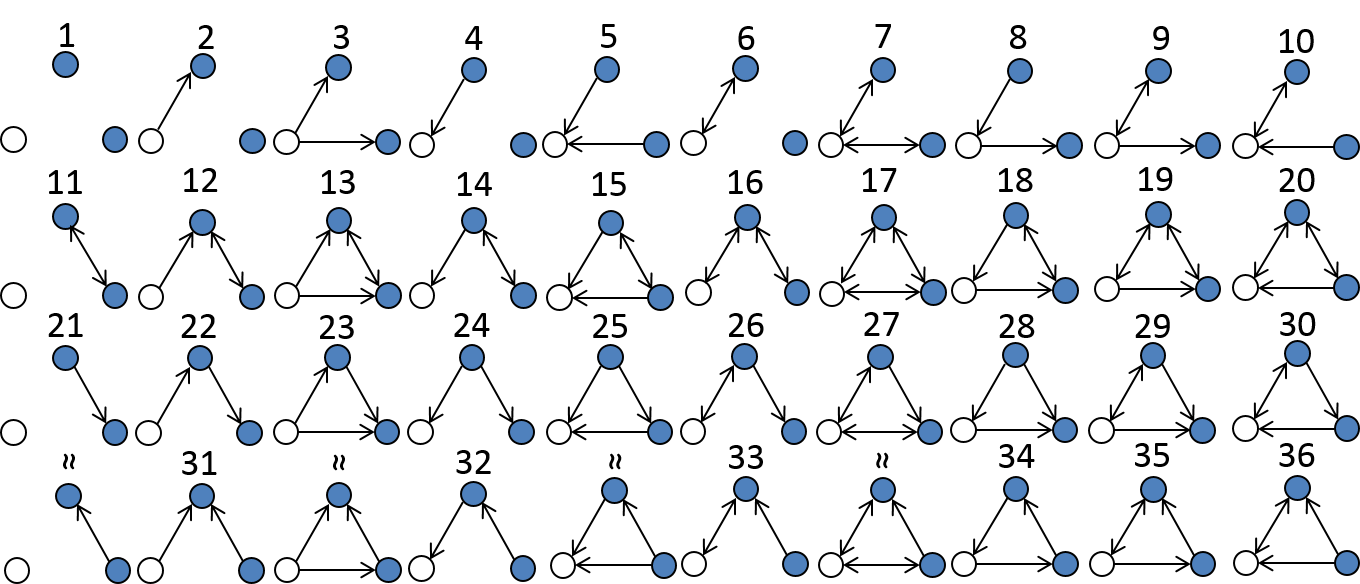}
\caption{Types of conditional triads}
\label{fig:triads}
\end{figure}

Figure~\ref{fig:triads} captures the 36 different ways an individual (white)
can be oriented towards two alters (blue) 
within a triad~\cite{burt90}. 
These orientations are the set of all {\em conditional triads},
which are defined by the structure of the three way relation based on the position of an individual within it. For example, triads 6 and 11 
are structurally identical (having two null and one mutual tie). In 
triad 11, the white user is isolated whereas in triad 6 she
is connected to an alter.
The entire structure of an ego-network can thus be represented by the
number and different types of conditional triads it is composed of. The
{\em conditional triad census}~\cite{wasserman94,cason13} of an 
ego-network is defined as a 36-element vector whose $i^{th}$ 
component represents the proportion of type $i$ conditional triads it is 
composed of.

Searching for ego-networks whose conditional triad censuses are 
similar is expected to lead to a meaningful grouping of users 
into social roles. This is because each triad configuration 
represents a sociological factor about how 
a user interacts with others~\cite{caplow68}. For example, triad 
32 has a user on the receiving end of a chain of interactions. 
If these interactions represent the passage
of information or rumors, it implies that the alter in the middle
of the chain is capable of manipulating what becomes shared with the 
user and may not be trustworthy.  In triad 5, the user receives 
interactions from two alters but chooses not to 
reciprocate. Ego-networks largely composed of this triad suggests
that the user receives many interactions but, for possibly 
selfish reasons, seldom chooses to reciprocate. By summarizing 
how frequently each of these triads appear, a conditional 
triad censuses succinctly models the strength of the different 
kinds of social factors that surround the nature of one's
interactions with others. These factors, taken together by
considering the entire census as a vector, therefore represents the 
circumstances and reasons why a user participates in a 
social system.

The number of and kinds of roles that exist in a social system can thus
be identified by:~(i) computing the conditional triad census of every
user; and~(ii) clustering users into groups based on the similarity 
(vector distance) of the conditional triad censuses. This approach is somewhat
related to discovering social groups in networks by searching for 
ego-networks that participate in similarly shaped $k$-cliques~\cite{hanneman05}
or -cores (sub-graphs where all nodes are connected to at least $k$ 
others~\cite{dorogovtsev06})~\cite{jamali06,rothenberg98,labatut12}.
However, searching for ego-networks that satisfy these strict 
requirements will only
identify sets of nodes surrounded by a similarly dense network and leave
hidden other nodes whose ego-networks are less connected but 
still have similar connectivity patterns. 
Such analysis also pays no consideration to the social forces or actions that 
drive users in cliques or cores to interact with each other, since the types 
of triads within the groups are ignored. Furthermore, it is difficult 
to know a priori what kinds of $k$-cliques and -cores correspond to relevant
social roles in a large-scale social system.
In comparison, the proposed approach learns significant 
structural patterns of ego-networks based on a feature reflecting
the types of social forces that bind a user and her connections
together. It leads to a classification where users in the same group 
participate and interact with their contacts under similar social circumstances
and forces, which speaks very closely to the notion of a social role€™.

\subsection{Dataset description}
The methodology is demonstrated by discovering social roles in two
popular online social systems, 
namely Facebook and Wikipedia. These systems were chosen because
they each serve a different purpose and provide distinct mechanisms 
for users to interact with each other. 
Facebook is used as a platform to informally share 
personal information, photos, and events with friends and family. 
Its interaction network is built 
by placing a directed edge from user $a$ to $b$
if $a$ posts at least one message on the wall 
(a collection of public messages)
of $b$. Wikipedia is an online encyclopedia with articles that are written and 
edited by an open community. Interactions on 
Wikipedia are defined by the modification of 
content contributed by another user; a directed edge 
from $a$ to $b$ is added if $a$ edited the text, 
reverted a change, or voted on approving an action to an 
article made by $b$. Both the Facebook and Wikipedia networks 
were constructed from 
publicly available datasets~\cite{viswanath09,maniu11}.
These datasets
only record the act of an interaction; it does not include any information
about the content of or the type of the interaction. Although the Facebook
data set is dated (interactions were recorded in 2009), privacy
improvements made to the Facebook API since make it all but impossible to
capture such interactions at scale today.

\begin{table*}
\centering
\begin{tabular}{c | c  c  c  c  c  c  }
Network & $|V|$ & $|E|$ & $\bar{d}$ & $\bar{C}$ & $\alpha_{in}$ & $\alpha_{out}$ \\ \hline
Facebook & 46,952 & 264,004 & 37.36 & 0.085 & 1.61 ($p > 0.732$) & 1.68 ($p > 0.964$) \\
Wikipedia & 138,592 & 740,397 & 10.68 & 0.038 & 1.54 ($p > 0.999$) & 1.83 ($p > 0.999$) \\ 
\end{tabular}
\caption{Dataset summary statistics}
\label{tab:summary}  
\end{table*}

Table~\ref{tab:summary} presents summary statistics 
for these interaction networks, illustrating how 
they vary in size, shape, and user behaviors. The relatively
small size (46,952 users, 264,004 pairwise interactions) of the Facebook network is due to the fact that it only represents
users within a single regional network (the Facebook social graph was divided
by user regions in its earliest form). The data also only represents user's whose accounts
were shared publicly, which was the default Facebook setting during the data collection 
period~\cite{viswanath09}. Despite its size and limit to a single regional network, 
previous work showed that all regional Facebook networks exhibited a similar structure (average path
lengths and diameter) and shape (clustering coefficients and assortativity)~\cite{wilson09}. More recent
studies further confirm that the structure and shape of these regional networks are very similar to the
structure of the modern global Facebook network~\cite{wilson12, ugander11};
therefore this data set is expected to contain similar interaction patterns as seen in the global Facebook network. The Wikipedia network is almost
three times the size of Facebook, with 138,592 users 
and 740,397 distinct pairwise interactions, 
but its clustering coefficient $\bar{C}$
is approximately 55\% smaller. These measurements 
suggest that Facebook users have a greater tendency to surround 
themselves within denser ego-networks compared to
Wikipedia users. The lower clustering coefficient of
Wikipedia could be explained by users who generally
limit themsleves to modifying
articles written by a specific group (perhaps representing a specific topic).

\begin{figure}
\centering
\begin{tabular}{cc}
  \includegraphics[width=55mm]{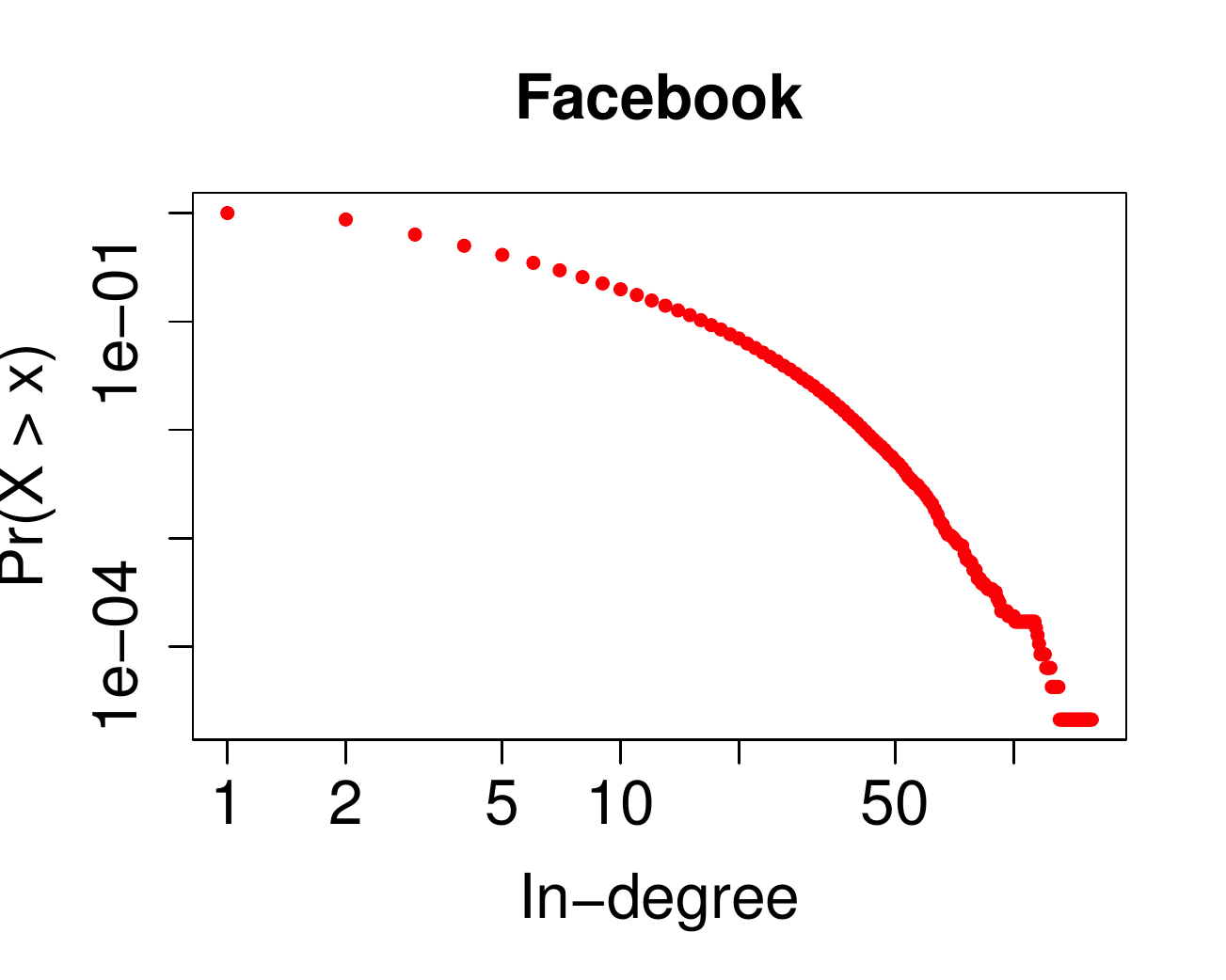} &  
  \includegraphics[width=55mm]{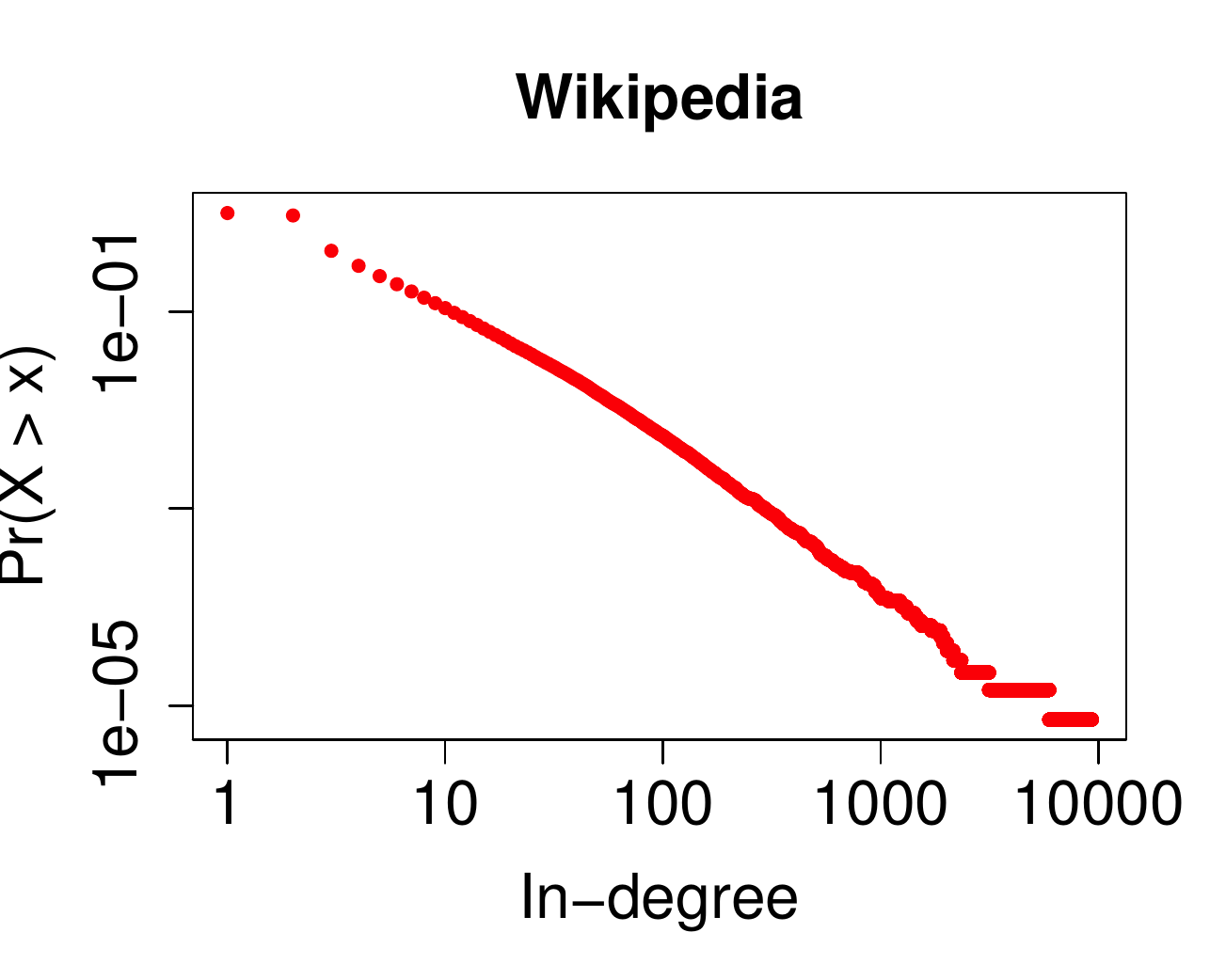} \\
  \includegraphics[width=55mm]{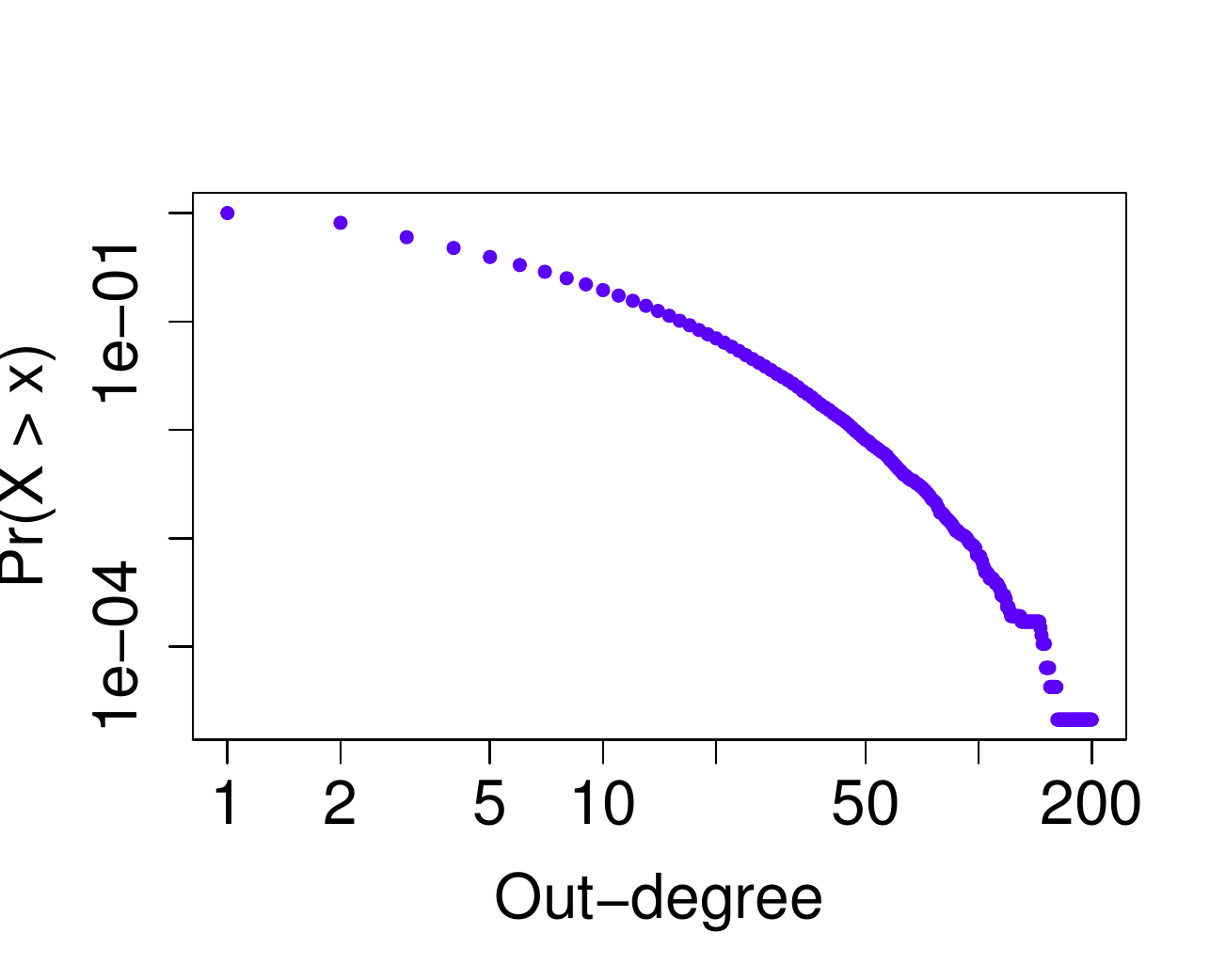} &
  \includegraphics[width=55mm]{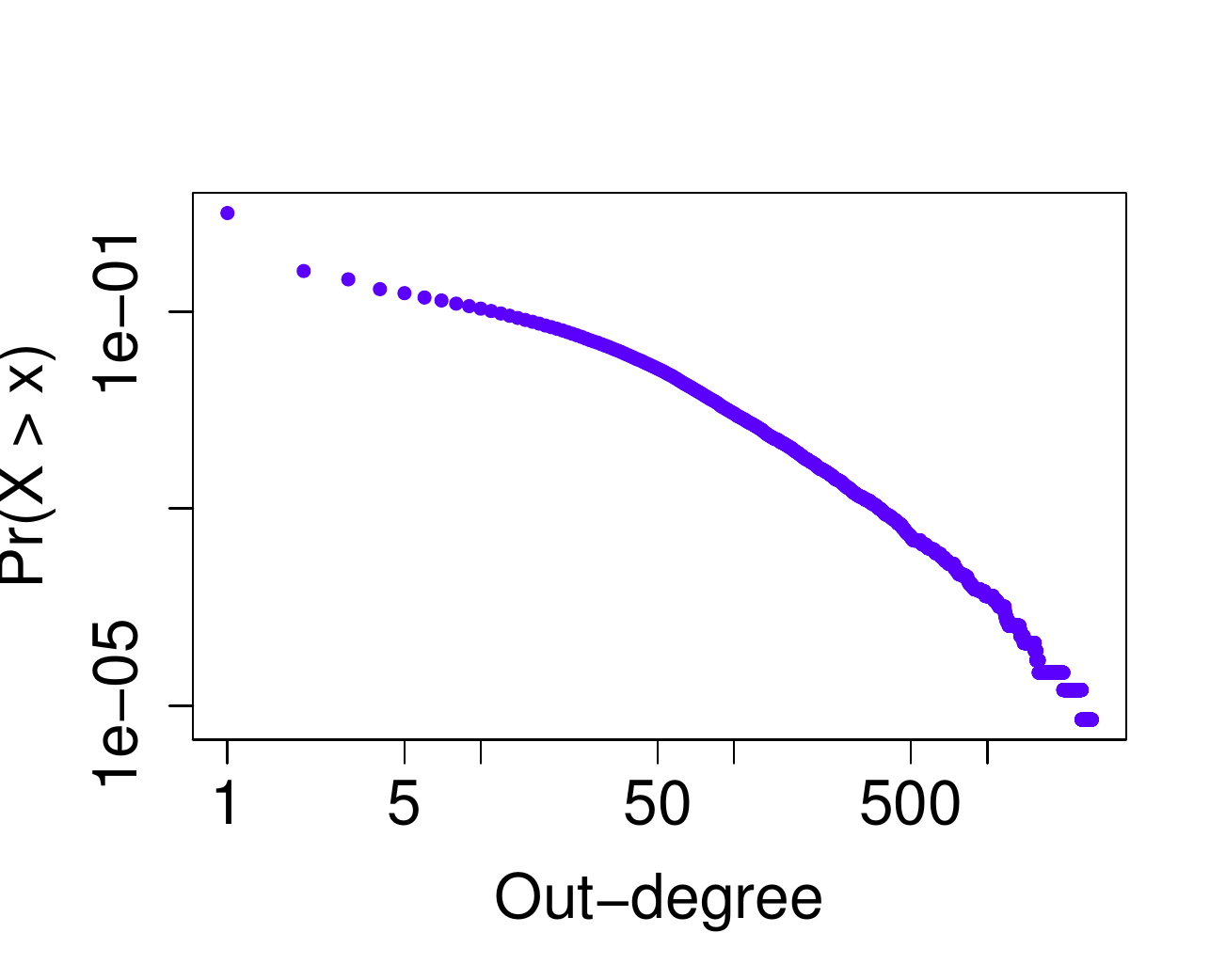} \\
\end{tabular}
\caption{In- (top) and out-degree (bottom) distributions}
  \label{fig:dist}
\end{figure}

The in- and out-degree distributions of each network is
presented in Figure~\ref{fig:dist}, which exhibit power-tailed shapes. 
The existence of power-law behavior is tested by a
maximum likelihood approach~\cite{clauset09}
and the resulting power-law exponents $\alpha_{in,out}$ are given
in Table~\ref{tab:summary}. The estimates of the power-law 
exponent are very reliable ($p > 0.95$; note that the test considers the hypothesis
$H_0$: the empirical data follows a power-tailed distribution) 
except for the in-degree
distribution of Facebook, which may be because its range only 
covers two orders of magnitude. 
A larger power-law exponent indicates that the distribution 
drops to zero faster in its right-tail~\cite{laqt}, 
hence the frequency with which users interact with others on Wikipedia
exhibits a smaller amount of variation compared to 
Facebook. In other words,
it is less likely to find a user who interacts with an 
unexpectedly high number
of others on Wikipedia compared to Facebook, and less likely
to find a user receiving many interactions from others
on Facebook compared to Wikipedia. 

\subsection{Network sampling}
Computing the conditional triad census of every ego-network 
requires an examination of $O(|V|^3)$ triples of 
users in an interaction network. This  
computational cost may be an insurmountable burden 
to compute conditional triad censuses in larger interaction networks
where the number of nodes are in the millions~\cite{ugander11}.
Furthermore, existing algorithms that can compute censuses in 
$O(|V|^2)$~\cite{moody98} or $O(|E|)$~\cite{batagelj01} only considers users'
{\em unconditional} triad censuses. An unconditional triad census is a 
16-element vector holding the proportion of all triads without regard
to the position of the user in her ego-network, making 
them incompatible with the proposed approach.
However, since the components of a 
conditional triad census are the {\em proportions} of triad types 
in an ego-network, the conditional censuses within a carefully selected
{\em sample} of the original network should be representative of the
conditional censuses in the original network.
A sample of a network $G$ is a new network $G_s = (V_s, E_s)$ 
where $V_s \subset V$, $E_s \subset E$, and $|V_s| = \phi|V|$ 
with $0 < \phi < 1$. 

A sampling method must ensure that the two critical local structural 
properties of ego-networks, namely the degree distribution and local
clustering coefficient distribution are preserved~\cite{holland78,fagiolo07}. For example, ego-networks with high degree 
will naturally tend to have triads with relations among multiple
alters, and lower (higher) cluster coefficients indicate a 
greater proportion of open (closed) triads.
However, na\"{\i}ve methods for network sampling do a poor job 
of preserving these local features.
A number of advanced sampling methods have been proposed, 
but each one can only preseve different types of structural features
of the full network~\cite{ahmed12}. Therefore, four
widely used graph sampling techniques for choosing
$V_s$ and $E_s$ were compared by their ability to 
preserve the degree distribution of the users' ego-network 
and their clustering coefficients. The techniques and their 
process are:

\begin{enumerate}
\item {\bf Vertex Sampling (VS)}: Let $V_s$ be a random sample of $\phi|V|$ vertices from
$V$ and define $E_s$ to be the set of all edges among the vertices in $V_s$ from $G$.
\item {\bf Edge Sampling (ES)}: Randomly choose an edge $e = (v_1,v_2)$ 
from $E$, add it to $E_s$, and add $v_1$ and $v_2$ to $V_s$ if they have not yet
been added. Continue to choose edges from $E$ until $|V_s| = \phi|V|$.
\item {\bf Forest Fire Sampling (FFS)}~\cite{leskovec06}: Choose a random vertex $v$ from $V$,
randomly select $p/(1-p)$ of its outgoing edges, and add theses edges to $E_s$.
Place every vertex incident to those added to $E_s$ into a set $V_*$ of `burned vertices'
and update $V_s$ by $V_s = V_s \cup V_*$. 
Randomly choose a burned vertex from $V_*$, and recursively repeat this process 
until $|V_s| = \phi|V|$. The parameter assignment $p = 0.7$ is used based on the recommendation of the method's 
authors~\cite{leskovec06}.
\item {\bf ES-i (ESI)}~\cite{ahmed12}: Randomly choose an edge $e = (v_1,v_2)$ 
from $E$ and add $v_1$ and $v_2$ to $V_s$ if they have not yet been added (note 
that $e$ is not added to $E_s$). Continue sampling until $|V_s| = \phi|V|$. Finally,
define $E_s$ to be the set of all edges among the vertices in $V_s$ from $G$.
\end{enumerate}

\begin{figure}
\centering
\begin{tabular}{cc}
  \includegraphics[width=65mm]{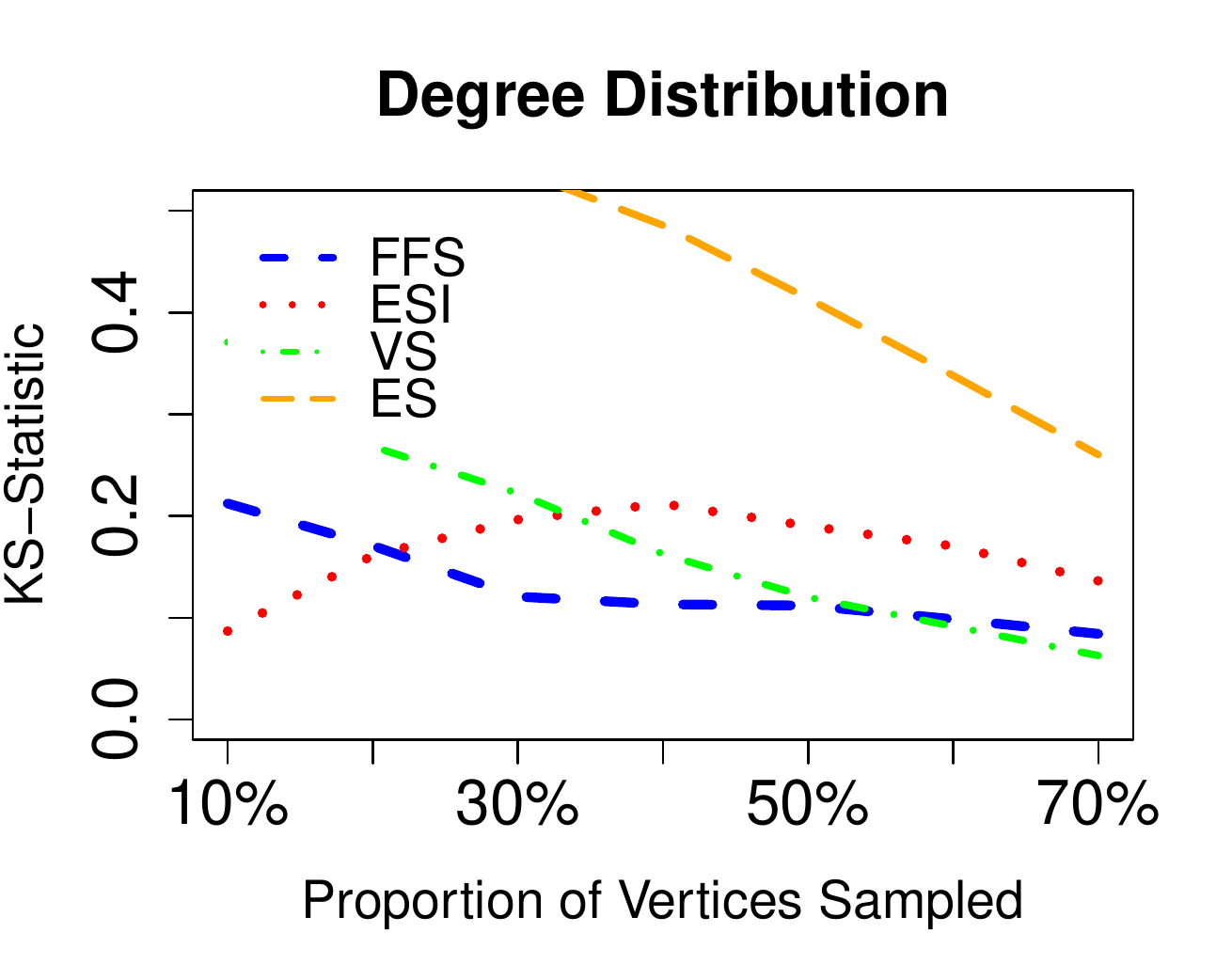} &
  \includegraphics[width=65mm]{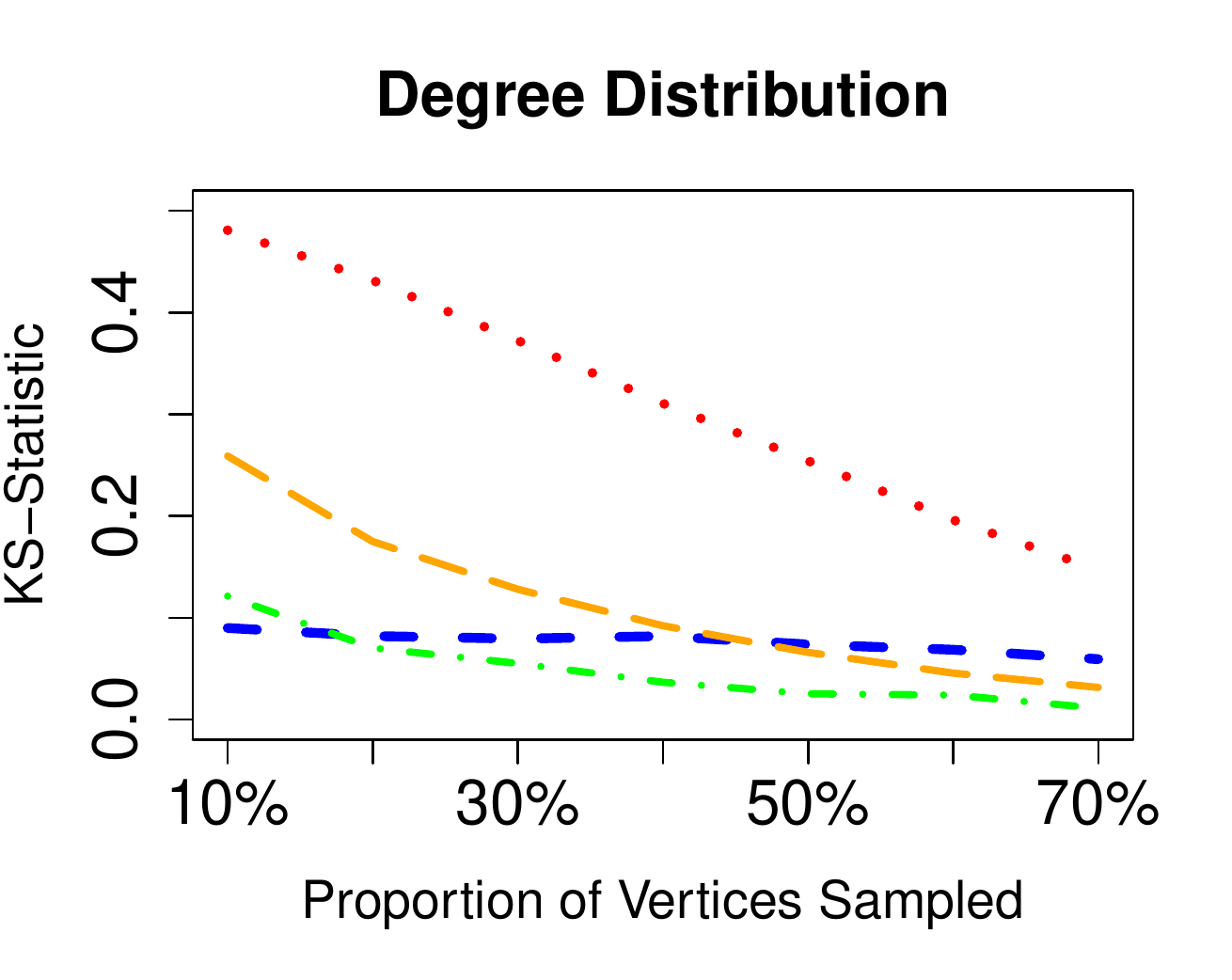} \\
  \includegraphics[width=65mm]{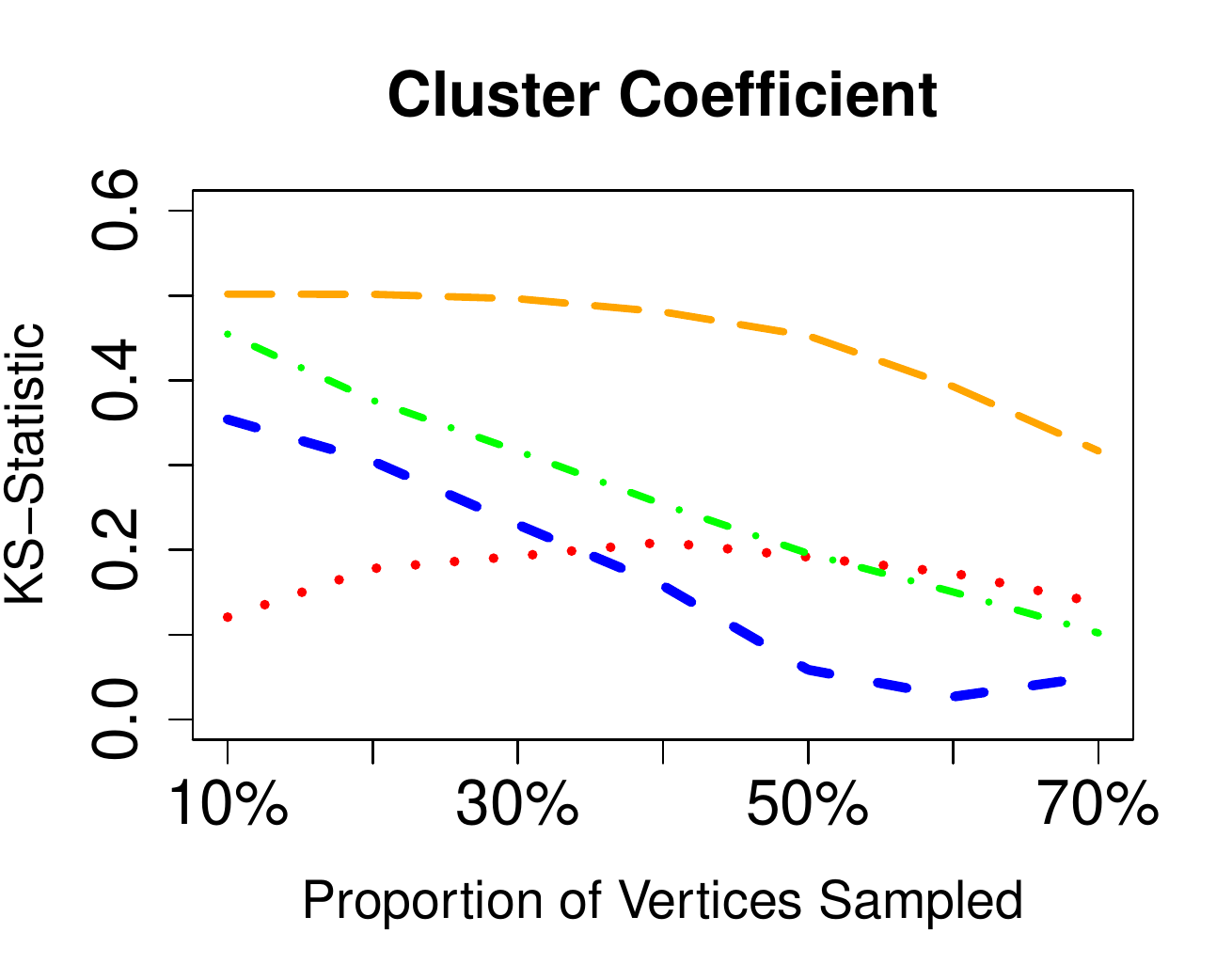} &
  \includegraphics[width=65mm]{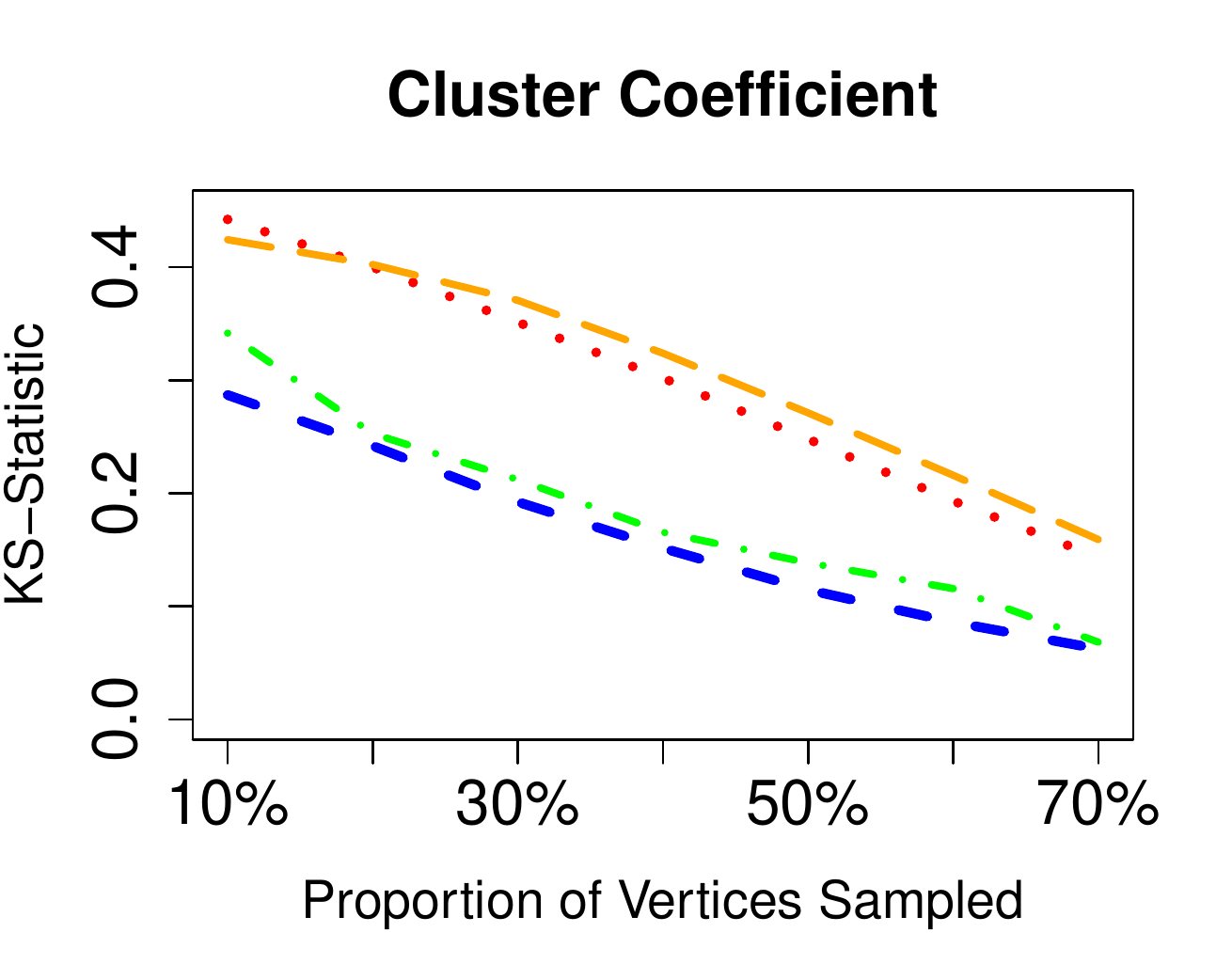} \\
    (a) Facebook & (b) Wikipedia \\ 
\end{tabular}
\caption{Comparison of graph sampling methods}
  \label{fig:ks}
\end{figure}

The Kolmogorov-Smirnov distance metric $D$ was used to compare how 
closely the degree and clustering coefficient distributions of 
samples $F_s$ taken with each method follow the distribution of 
the original network $F$. It is defined by the largest difference 
of a point taken from original distribution $F$ to the  
distribution of the sample $F_s$: $$D = \sup_x |F_s(x) - F(x)|$$
Figure~\ref{fig:ks} compares the average $D$ over $100$ 
samples taken by each method for different values of $\phi$. 
For the Facebook network, FFS does the best job ($D < 0.2$) at preserving 
both degree 
and clustering coefficient distributions for modest sample sizes
($\phi \geq 0.33$). FFS samples of the Wikipedia network best preserves the 
clustering coefficient distribution for any sample size and
$0.2 < \phi < 0.3$ FFS and VS are similarly faithful 
to the original network's degree distribution. Ultimately, FFS sampling
is found to be able to preserve the local structure of both networks 
even for small sample sizes. 

\begin{figure}
\centering
\begin{tabular} {cc}
  \includegraphics[width=55mm]{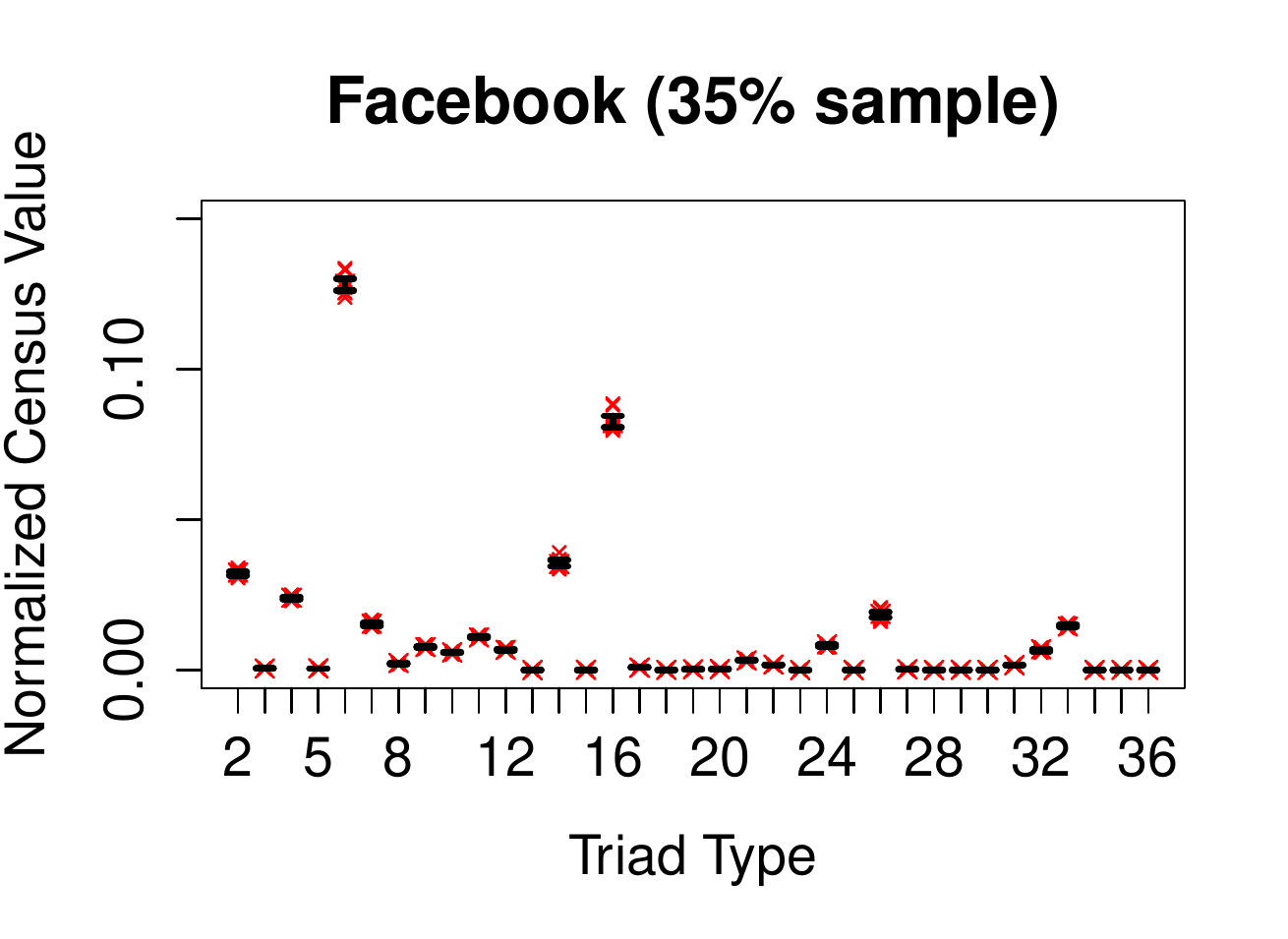} &
  \includegraphics[width=55mm]{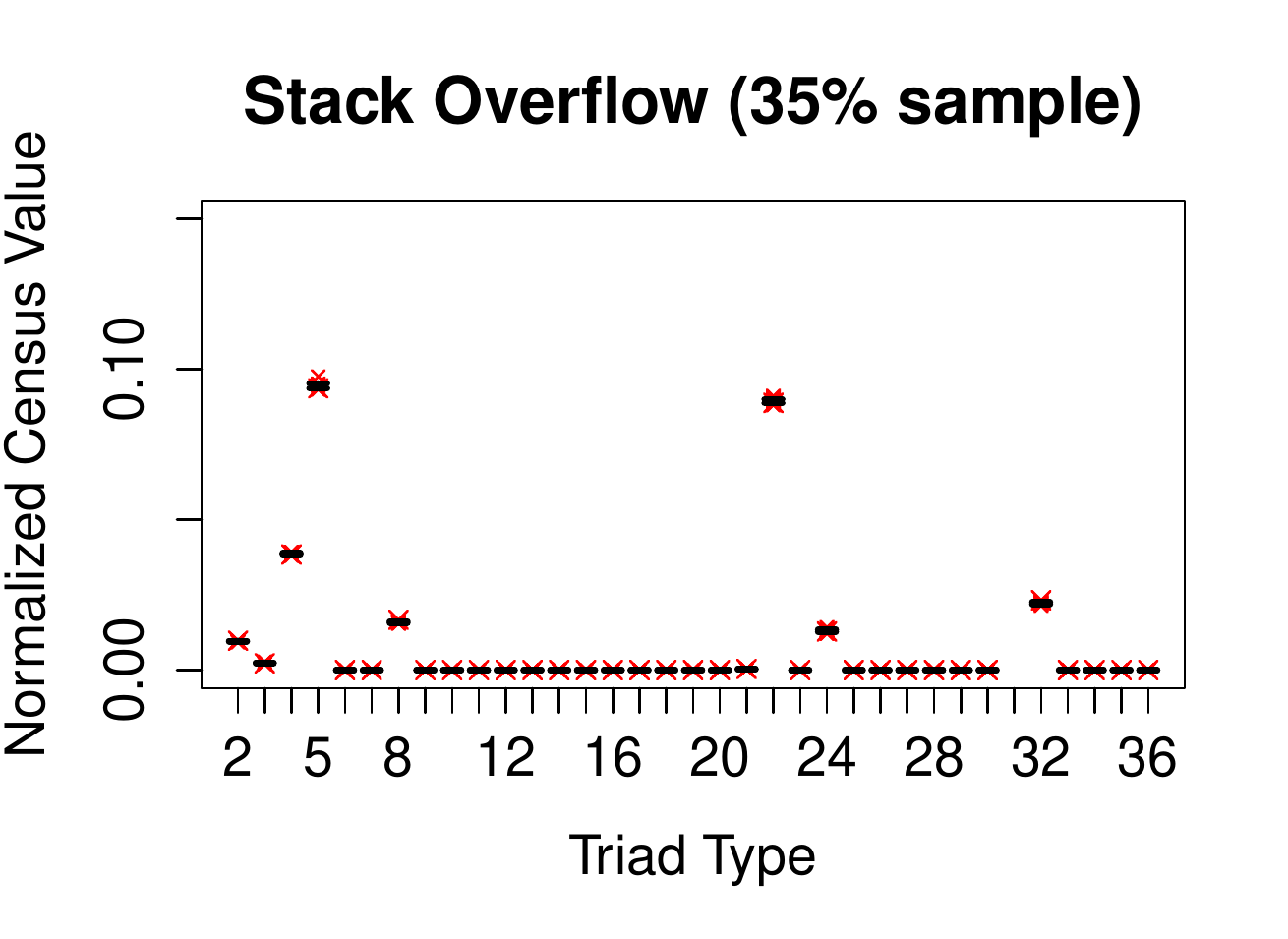} \\
  (a) Facebook & (b) Wikipedia \\
\end{tabular}
\caption{Triad Role Census Sample Values with 95\% Confidence Intervals}
  \label{fig:samples}
\end{figure}

A value of $\phi$ that provided a reasonable trade-off
between computational speed and sample consistency was searched
for. Figure~\ref{fig:samples} plots the average value 
of each component of conditional triad censuses taken 
from $n=20$ independently generated FFS samples of each network for
$\phi = 0.35$ (triad 1 is excluded because of its disproportionately high
frequency) and the 95\% confidence interval of the proportions. 
The proportion of triad types across the samples are similar and feature 
small confidence intervals. Since the computation cost of computing
triad censuses at this sampling level is very reasonable
(less than 30 minutes in a parallel computation over
three cores of an Intel i5 processor), the setting $\phi = 0.35$ is used
for role analysis. 

\subsection{Census clustering}
$k$-means clustering, a common and flexible algorithm 
for discovering latent groups in 
data~\cite{docCluster,videoCluster,yeastcluster}, is used to 
separate users into roles. 
$k$-means clustering defines $k$ centroid positions in 
the vector space and assigns each conditional triad census (and hence user) 
to a cluster based on the centroid it is most similar to.
Since the components of the censuses take a 
value between $0$ and $1$, this similarity is defined as 
the $\ell_2$-norm of their difference vector.
After the assignment of conditional triad censuses to clusters, the
position of the centroid of each cluster is updated. Censuses are then
reassigned to their closest centroid, and the process repeats 
until there are no changes to any cluster assignments.

\subsubsection{Dimensionality reduction}
\begin{figure}
\centering
\begin{tabular}{cc}
  \includegraphics[width=55mm]{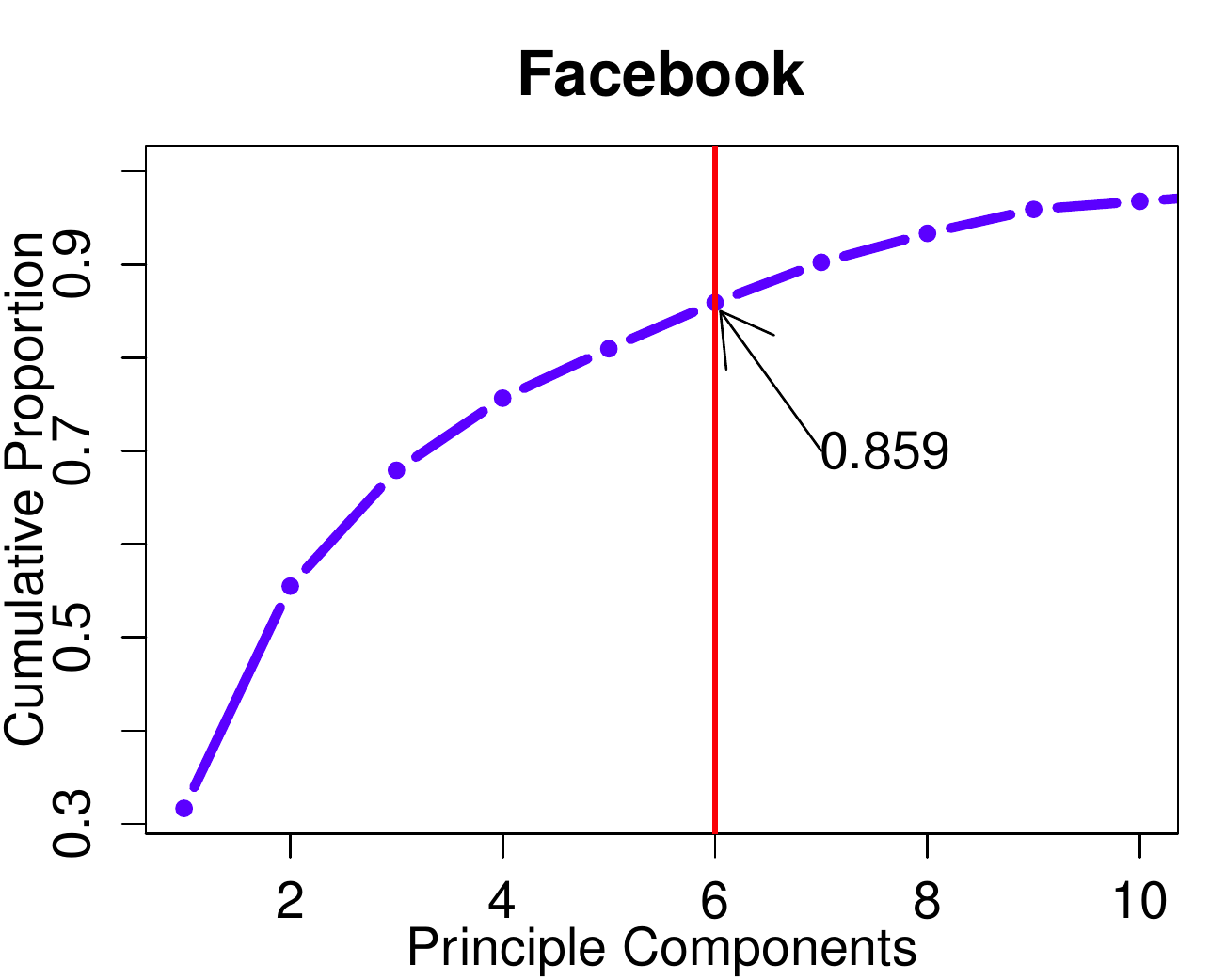} &
  \includegraphics[width=55mm]{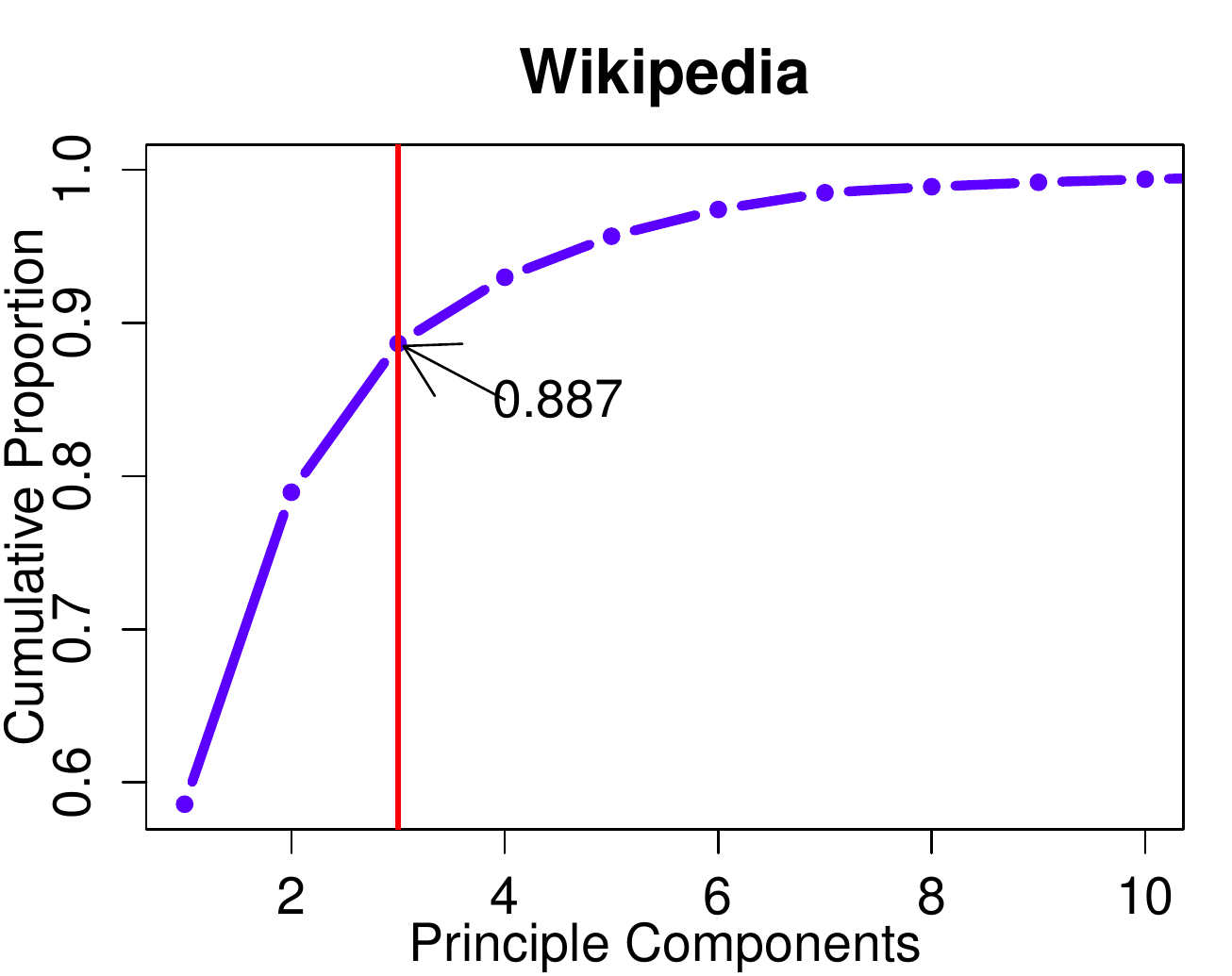} \\
  (a) Facebook & (b) Wikipedia \\ 
\end{tabular}
\caption{Scree plots}
  \label{fig:scree}
\end{figure}

Figure~\ref{fig:samples} indicates that many components of 
the conditional triad censuses are close to or equal to $0$. A 
dimensionality reduction technique, namely principle component
analysis (PCA)~\cite{pca}, is therefore applied to the conditional triad
 censuses. PCA identifies a projection of the data into a lower 
dimensional subspace that preserves as much variation within the 
original space as possible. Figure~\ref{fig:scree} plots 
the proportion of variation within the original dataset that 
is retained when we use PCA to reduce the data into smaller numbers 
of principle components. The smallest dimensional space that still
preserved a large proportion of the variation in the data
($>$ 85\%) was chosen, as indicated by the red line in Figure~\ref{fig:scree}.
The figure suggests that PCA finds a significantly lower 
dimensional space for clustering the 
conditional triad census of every network, from 36 dimensions to just
6 and 3 for the Facebook and Wikipedia interaction networks 
respectively. 

\subsubsection{Clustering evaluation}
\begin{figure}
\centering
\begin{tabular}{ccc}
  \includegraphics[width=40mm]{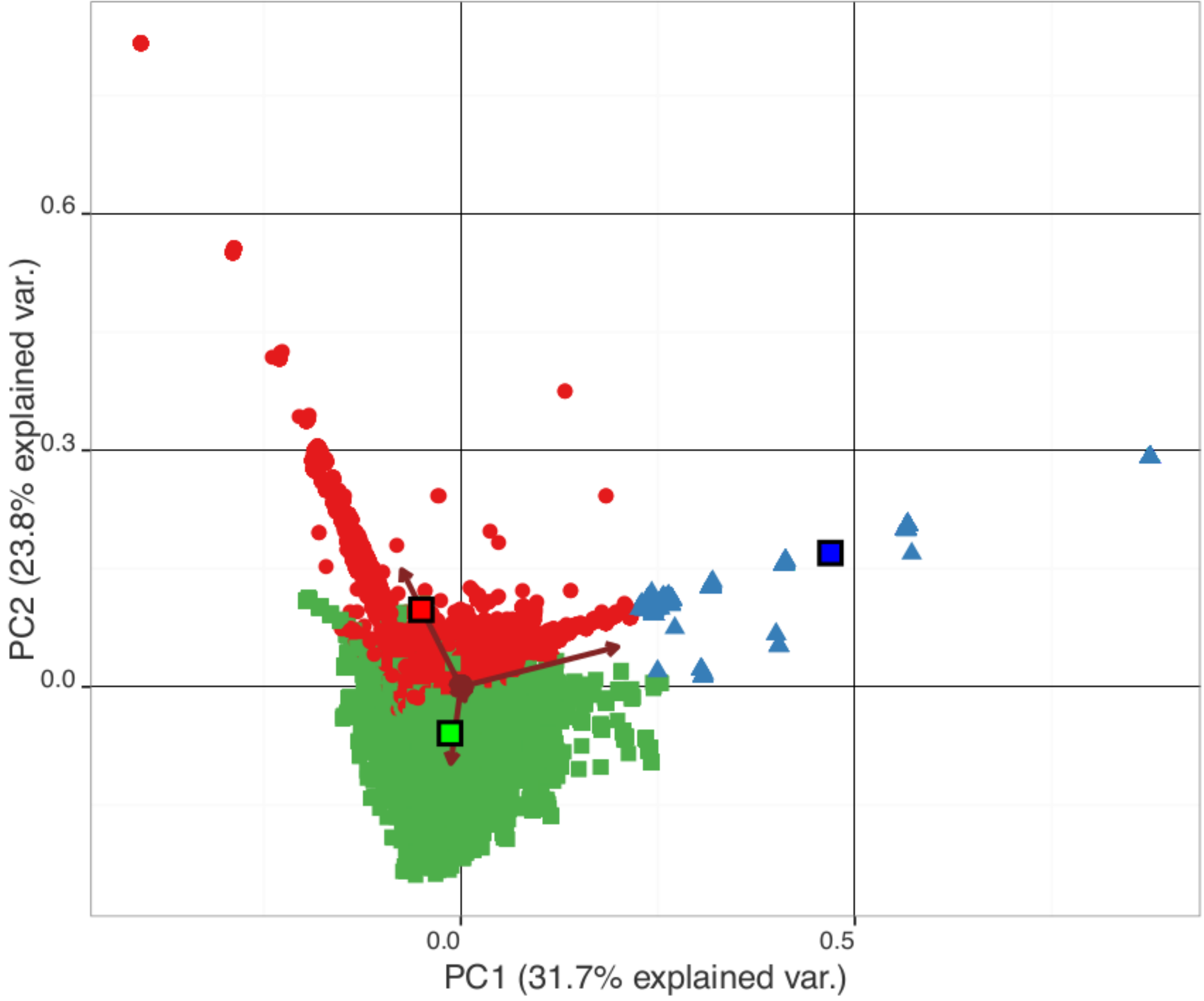} &
  \includegraphics[width=60mm]{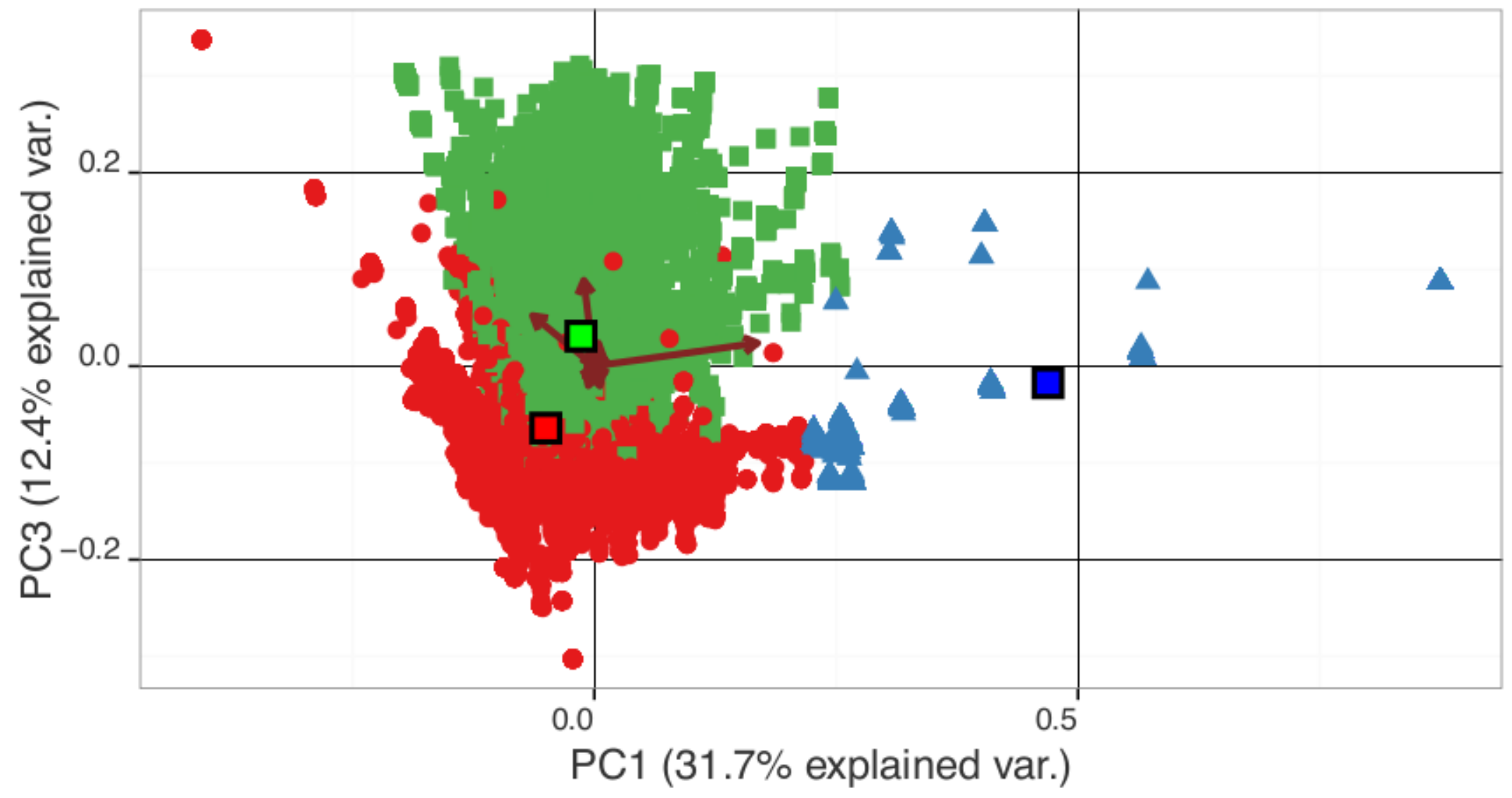} &
  \includegraphics[width=50mm]{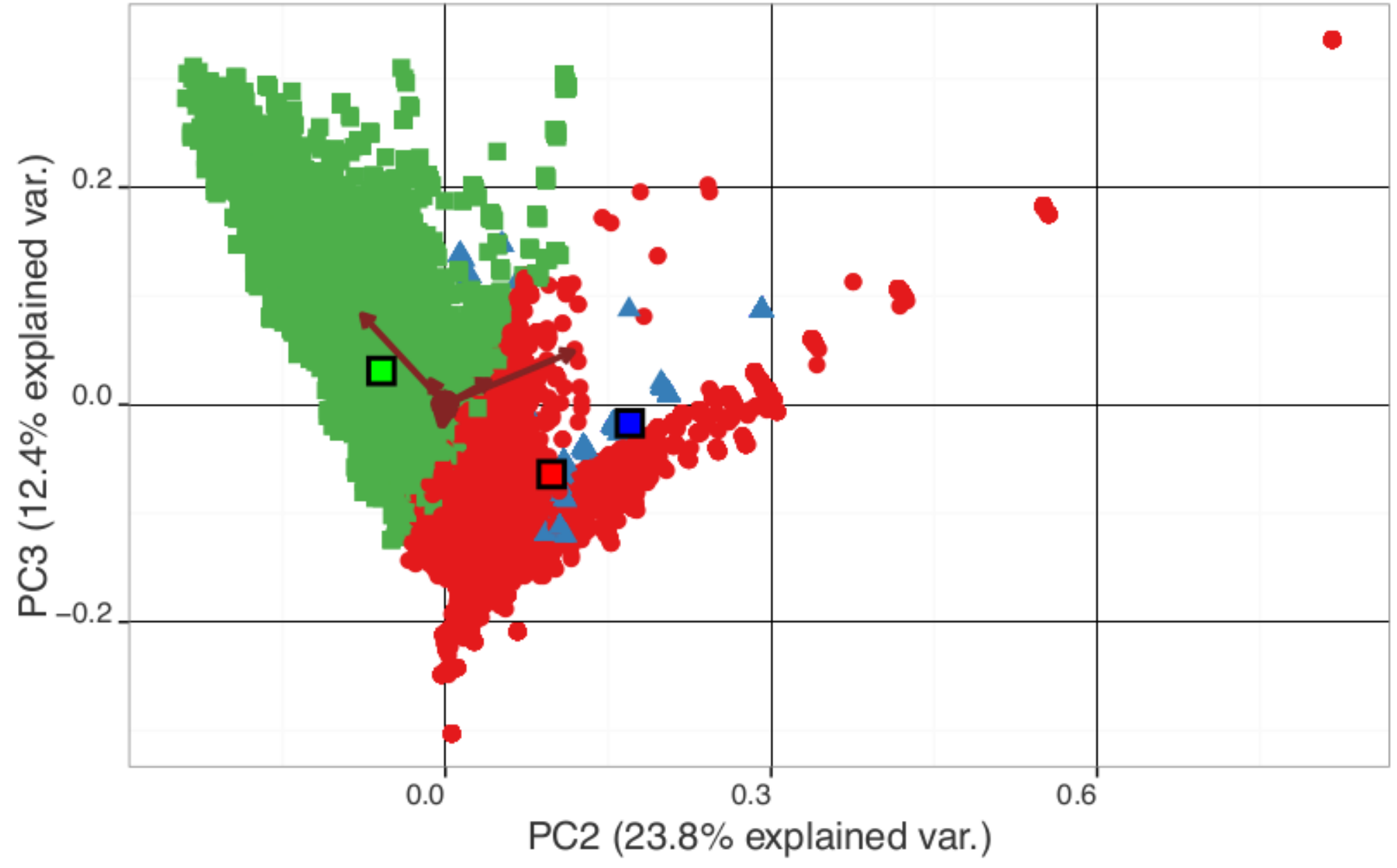}  \\

  \includegraphics[height=45mm]{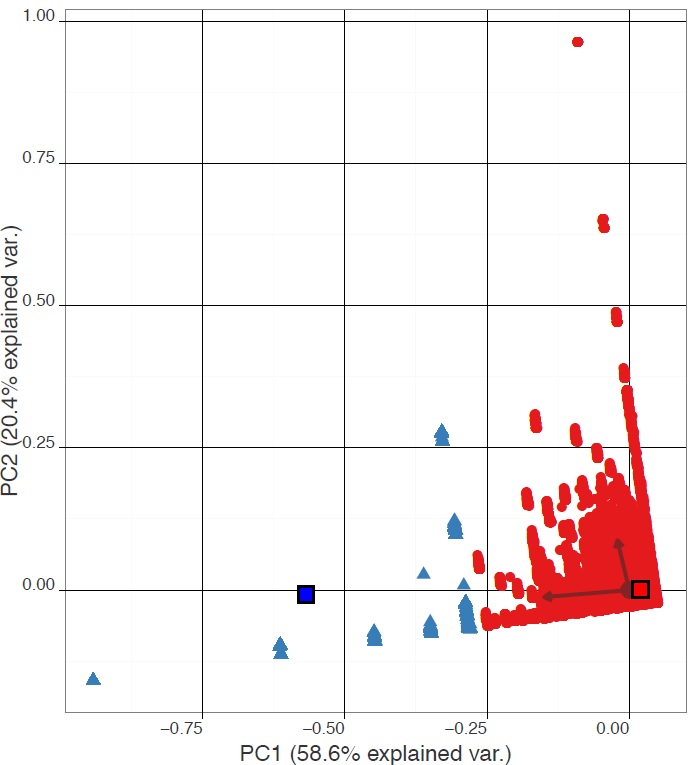} & 
  \includegraphics[height=45mm]{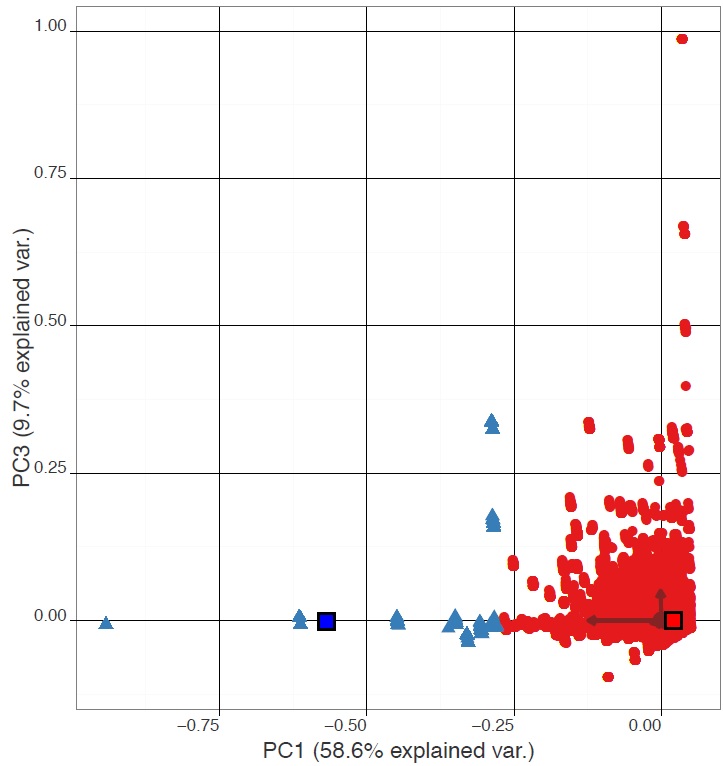} & 
  \includegraphics[height=45mm]{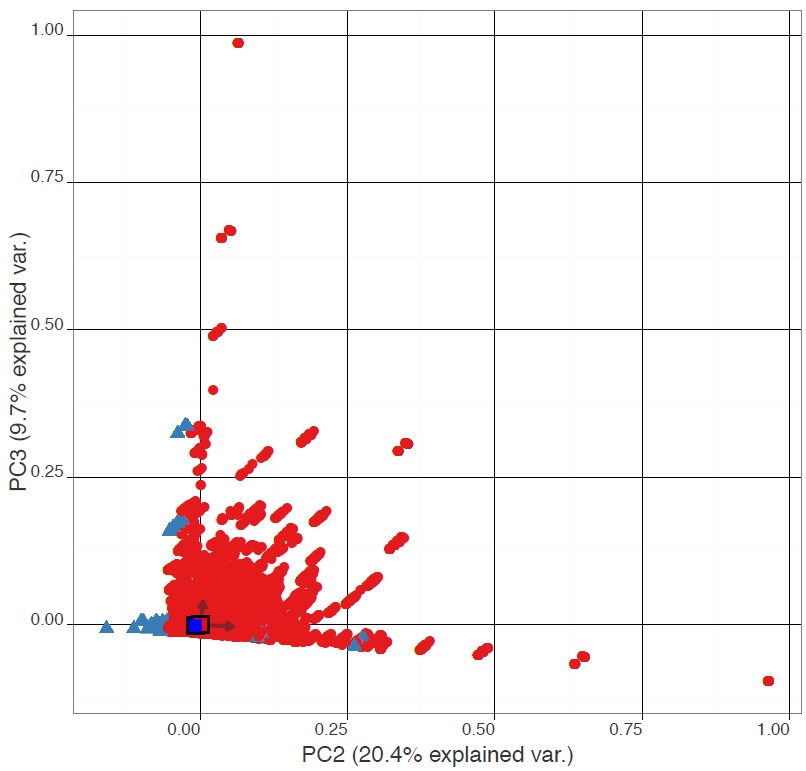} \\
    (a) Principle Comp. 1 and 2 & (b) Principle Comp. 1 and 3 & 
  (c) Principle Comp. 2 and 3 \\
\end{tabular}
\caption{Clusters along the first three principle components for 
Facebook (top) and Wikipedia (bottom)}
  \label{fig:clusters}
\end{figure}

$k$-means clustering requires the number of 
clusters $k$ to divide the data into to be chosen beforehand, forcing
an analyst to assert the specific number of social roles that
may exist in the system. Instead, the  
silhouette coefficient metric~\cite{Tan06} $SC_{\hat{C}^k}$ 
is used to  quantitatively evaluate the quality of clusters
for different values of $k$, so that the $k$ yielding the `best' 
clustering is chosen. It is defined as follows: consider a 
division of censuses into $k$ clusters $\hat{C}^k = \{C_1,C_2,...C_k\}$. Let
$\alpha(\x) = d(\x,C^*_i), x \in C_i$ be the distance from the 
vector $\x$ to the centroid $C^*_i$ of its assigned cluster $C_i$
(measuring intra-cluster distance) and 
$\beta(\x) = \min_{C_j \in \hat{C}^k, C_j \neq C_i} d(\x,C^*_j)$
be the distance from $\x$ to the centroid of the nearest cluster $C_j$
$\x$ is not assigned to (measuring inter-cluster distance). The silhouette of $\x$ is defined as: 
$$\phi(\x) = \frac{\beta(\x)-\alpha(\x)}{ \max(\beta(\x),\alpha(\x))}$$
Note that $\phi(\x)$ approaches $1$ as the separation between the 
cluster $\x$ is assigned to and the nearest other cluster 
increases. The average silhouette of every clustered vector defines
the silhouette coefficient of a clustering $\hat{C}^k$:

\begin{figure}
\centering
\begin{tabular}{cc}
  \includegraphics[width=55mm]{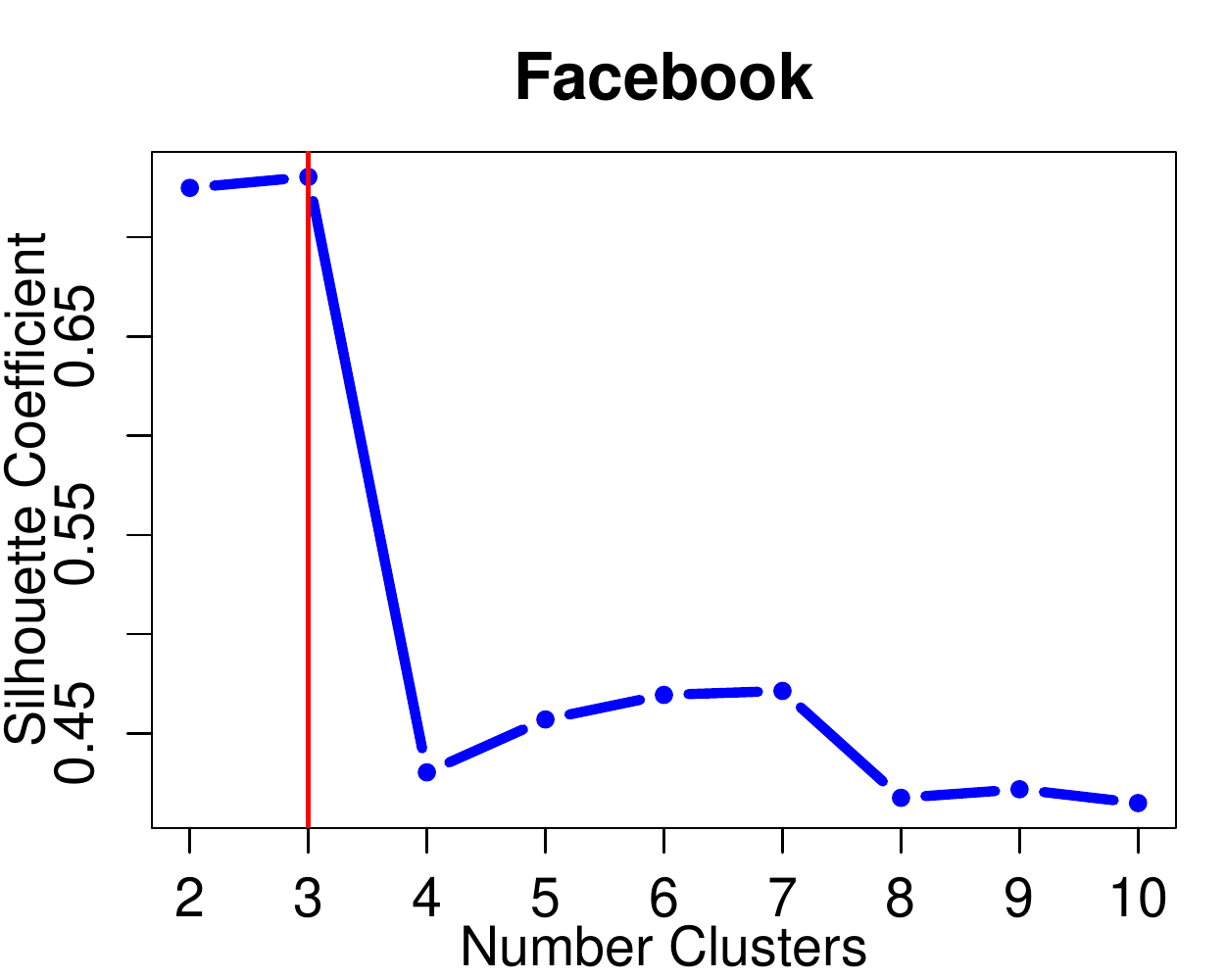} &
  \includegraphics[width=55mm]{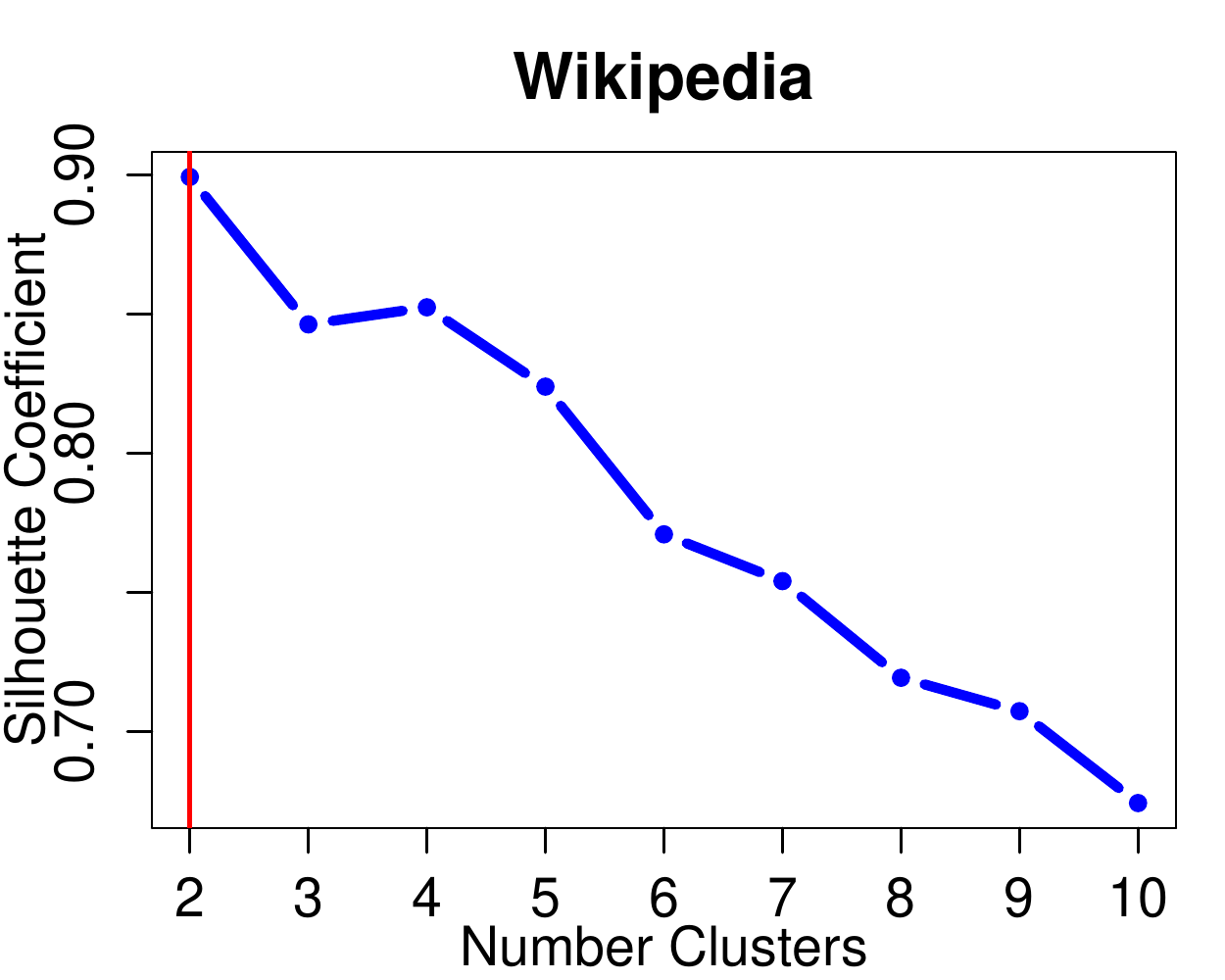} \\
  (a) Facebook & (b) Wikipedia \\
\end{tabular}
\caption{Silhouette coefficients}
  \label{fig:silco}
\end{figure}

$$SC_{\hat{C}^k} = \frac{\sum_{\x \in \X}\phi(\x)}{|\X|}$$
where $\X$ is the set of all data vectors. Previous studies indicate that
values of $SC_{\hat{C}^k}$ greater than 0.7 means the algorithm achieved 
superior separation, and values between 
0.5 and 0.7 indicate a reasonable separation~\cite{Tan06}.
For a given value of $k$, we ran 50 $k$-means clusterings 
over the PCA-reduced conditional triad censuses using 
different random initializations of the centroid positions. 
Figure~\ref{fig:silco} plots the average $SC_{\hat{C}^k}$ 
of these trials for $2 \leq k \leq 9$. It reveals
excellent clustering solutions at 
$k=3$ and $k=2$ clusters for the Facebook
and Wikipedia censuses, with silhouette coefficients of $0.73$ and $0.90$, 
respectively. A qualitative validation of the adequacy of a clustering 
solution is also given in Figure~\ref{fig:clusters}. Here, 
the conditional triad censuses
in a space defined by the first three 
principle components are assigned a marking 
and color corresponding to their cluster assignment. The Facebook clustering
solution, given in the top panels of the figure, 
discovers a role (the red cluster of circle points) that exhibits 
large variation along two principle components. In contrast, a second role 
(the green cluster of square points) varies strongly along the third component.
The smallest cluster (blue cluster of triangle points) only varies along
the first component. Since the clusters exhibit little variation along
different directions, different subsets of conditional triads must appear in 
similar proportions within the censuses of the same group. The Wikipedia 
clustering solution, given in the bottom panels of Figure~\ref{fig:clusters},
also finds that the two clusters vary along the direction of different principle components: the red cluster of circle points vary along the second
and third components, while the blue cluster of triangle points mainly varies along the first component. 


\section{Triad-based Role Analysis}
\label{sec:analysis}
In this section, the kinds of social roles that emerge
from our clustering analysis is analyzed. For this purpose, 
the average centroid positions $C^*_i$ over a clustering
result was identified and the user $u^*_i$ whose
conditional triad census is located closest to $C^*_i$ was found. 
$u^*_i$ is defined as the ``central user'' of role $i$ whose 
ego-network is the ``central structure" of the role. Due to its
position in the cluster, this ``central structure" 
represents the way a prototypical user having this role embeds 
herself within the social system. In other words, the ego-network structure
of users in role $i$ are most similar
to $C^*_i$ compared to any other central structure on the network. 
Each central structure is given a social role
label based on a subjective interpretation of the
user's position within it. The label captures
the way users of a role interact with others in the system,
and how the structure representing a role affects 
the kinds of interactions that are possible. The role labels may not
be applicable to all social systems, although it is feasible that 
systems created under
a similar context (e.g. social sharing sites) exhibit similar central role
structures and labels. The central role structures discovered in the Facebook and Wikipedia networks, and support for the emergence of these roles in the
literature, are presented next.

\subsection{Facebook}
\begin{figure}
\centering
\begin{tabular}{ccc}
  \includegraphics[width=55mm]{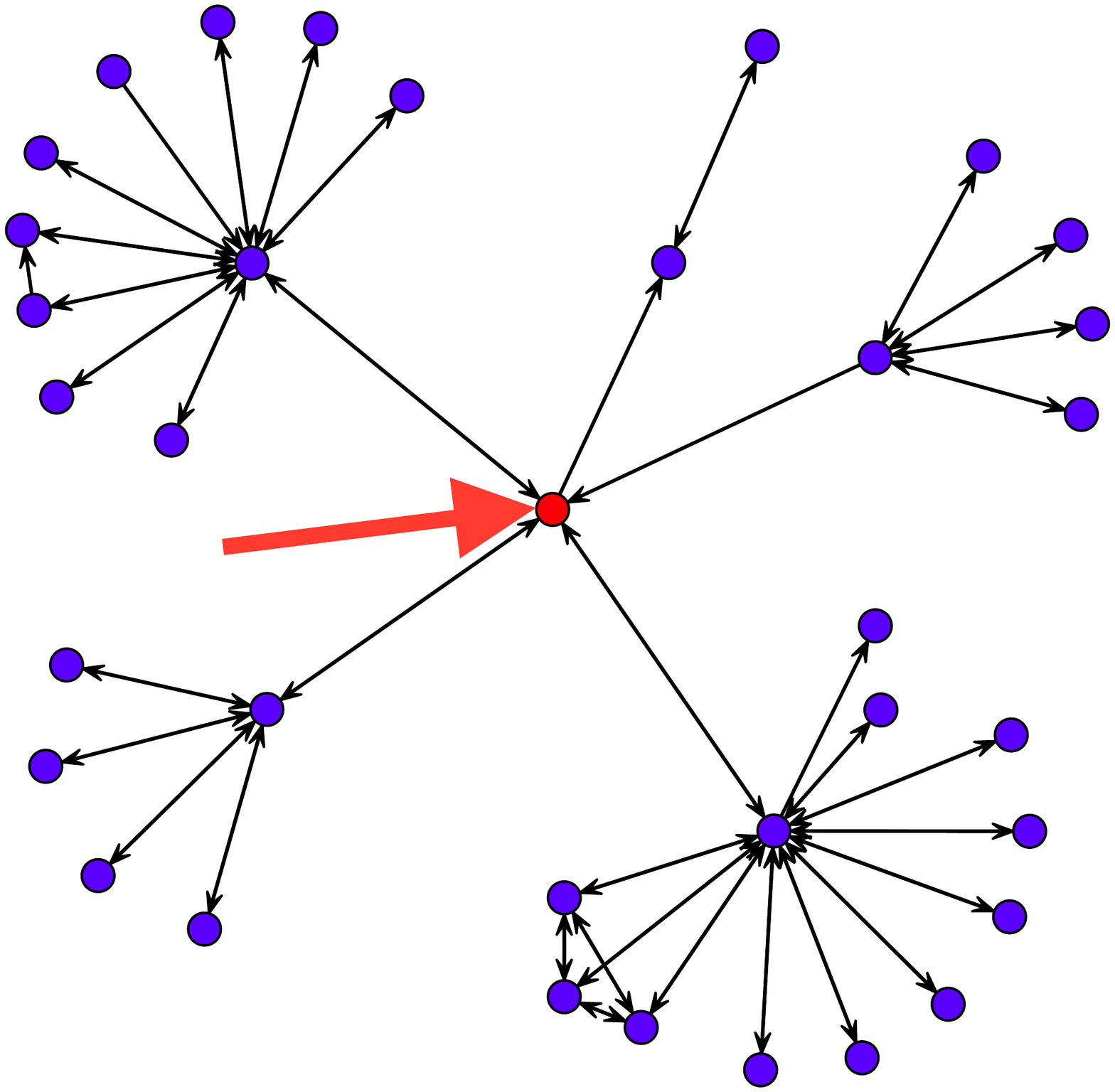} &
  \includegraphics[width=55mm]{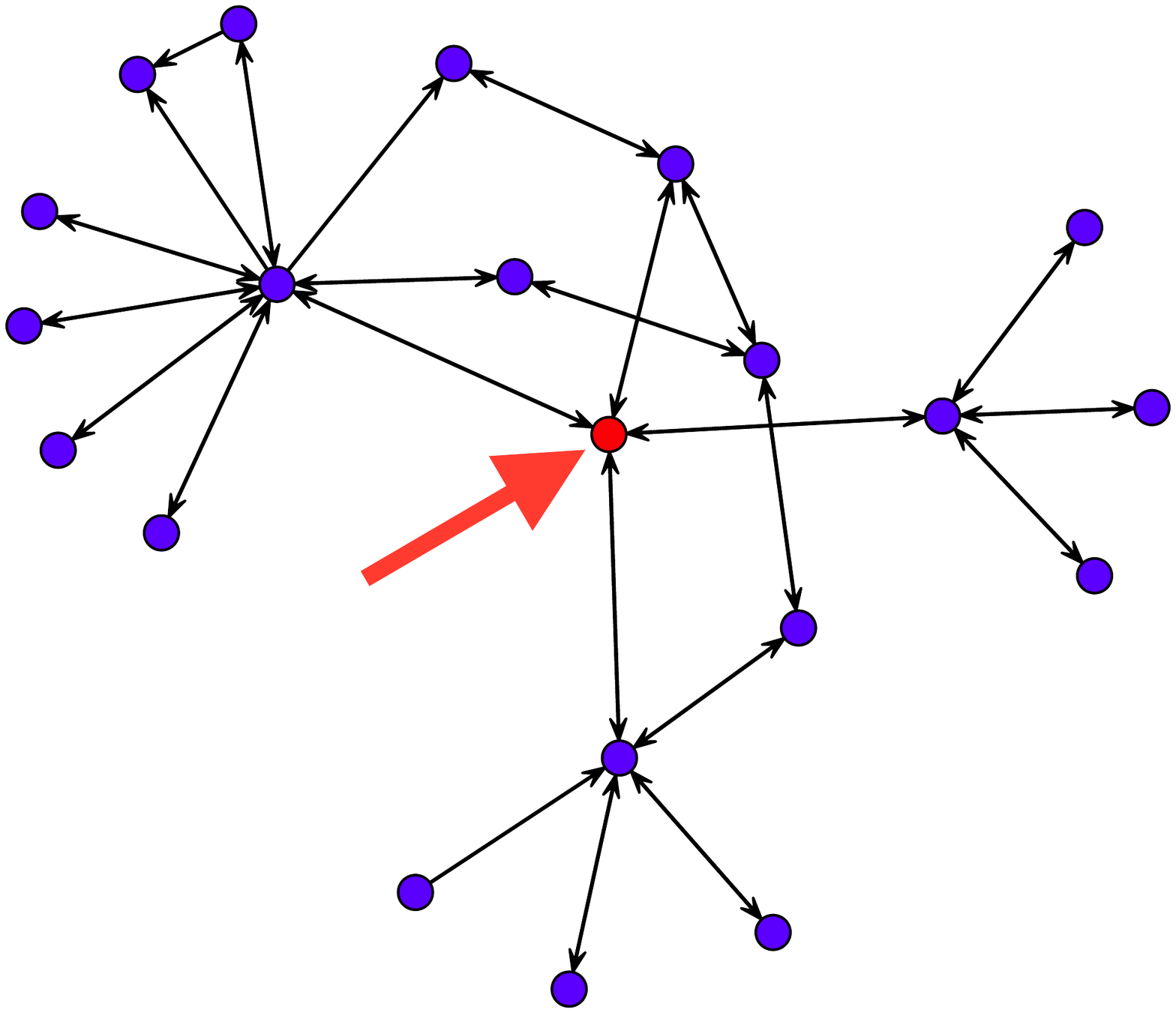} &
  \includegraphics[width=55mm]{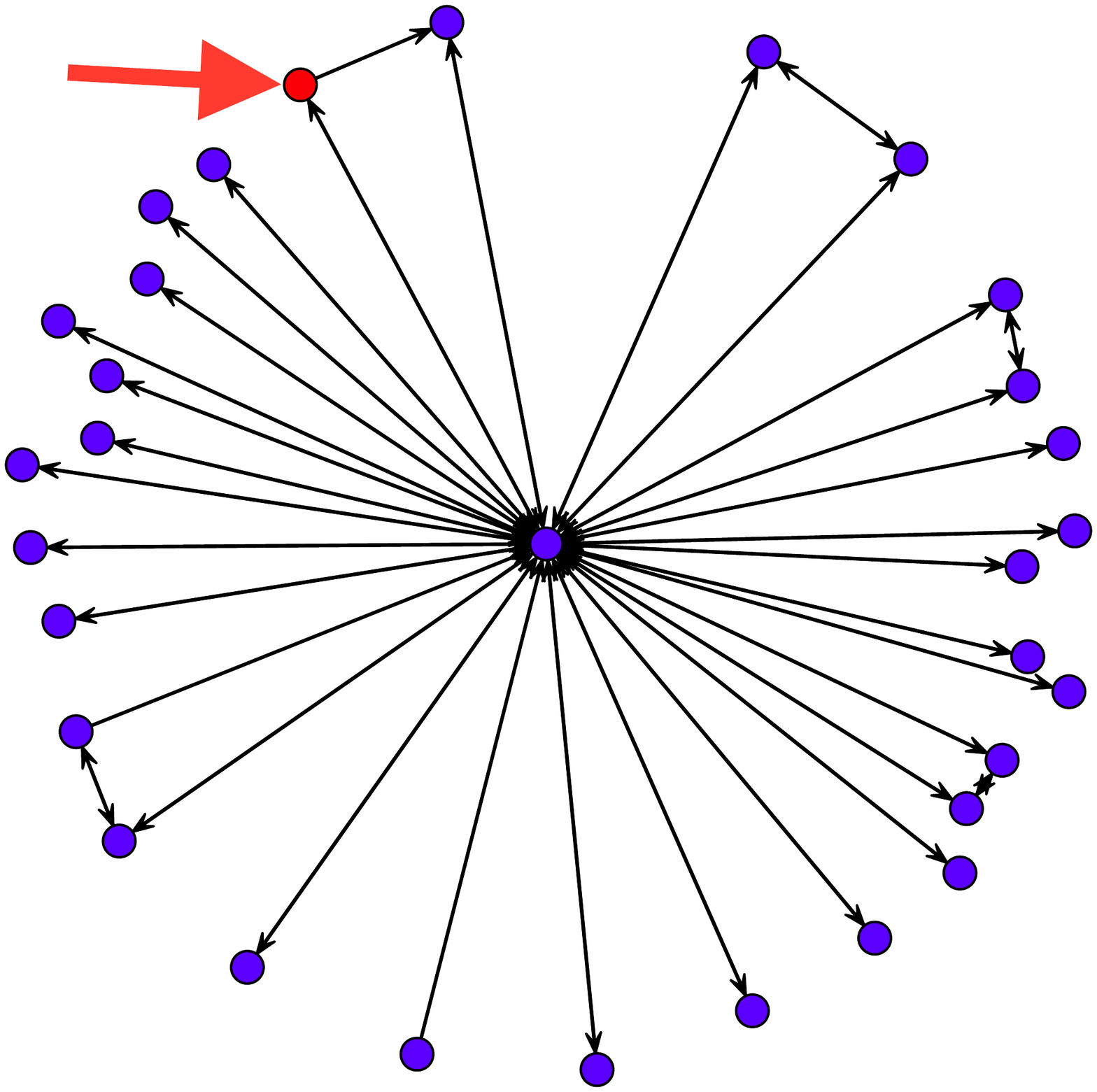}\\
    (a) & (b) & (c)  \\
\end{tabular}
\caption{Central role structures: Facebook}
  \label{fig:fbb}
\end{figure}

Figure~\ref{fig:fbb} presents the central role structures of the three social
roles found on Facebook. A label representing each role structure and the 
proportion of users falling under each are presented in Table~\ref{tab:fbb}. 
In these figures, the red node (with a red arrow pointing to it) 
corresponds to the central user and the blue nodes are the members 
of her ego-network.
The structure in Figure~\ref{fig:fbb}(a) represents a social role the 
majority of all Facebook users (56.6\%) fall into 
where a user is centrally embedded between many disconnected groups of others. 
She lies in a position critical for maintaining connectivity between 
communities, and hence, lies in the brokerage position of many 
open triads. These many open triads give users in this role
many opportunities to control if and how information exchanges from one 
group to another. However, given the fact that Facebook is used as a platform
for social sharing, such users may never decide to share information between
communities when they represent different social circles. For example, one can envision the user in Figure~\ref{fig:fbb}(a) to 
be sitting between groups that may correspond to colleagues at work, 
relatives, personal friends, and work colleagues. A user may never want
personal information shared among relatives to be revealed by 
work colleagues, and may want conversations, rumors, and other information shared
among friends to never be exposed to family members and work colleagues. 
That a majority of Facebook users fall into a social role that brokers among
many disconnected circles is not surprising; many past research studies 
have shown that most Facebook users face identity management and multiple
presentation issues while interacting on the site~\cite{dimicco07,skeels09,barkhuus10,lampinen09}. 
Identification of these ``social group managers" is thus a way of finding the bridges 
or weak ties~\cite{burke13} in the network based on structural patterns rooted in social theory.

\begin{table}
\centering
\begin{tabular}{c | c | c}
Role label & Structure & Proportion of users \\ \hline\hline
Social Group Manager & Figure~\ref{fig:fbb}(a) & 56.6\% \\
Exclusive Group Participant & Figure~\ref{fig:fbb}(b) & 28.4\% \\
Information Absorber & Figure~\ref{fig:fbb}(c) & 15.0\% \\
\end{tabular}
\caption{Facebook roles}
\label{tab:fbb}
\end{table}

28.4\% of Facebook users fall into the role represented 
by the central structure of Figure~\ref{fig:fbb}(b). This structure represents 
a user that has surrounded herself around a web of interactions running 
between her first-degree connections. This small percentage of users only participates in a single, tight-knit community of others rather than managing
many disconnected groups. Such a role may represent users who only choose
to `friend' and interact with a collection of others that share many mutual
connections, and does not need to manage multiple disconnected social circles.
Such patterns are known to be more prominent in the ego-networks of 
Facebook users
who are more willing to share information with many others, or does
not feel a need to consider identity management on the 
social network~\cite{gonzalez07,mcandrew12,pempek09,bosch09}. 
Such ``exclusive group participants" 
may therefore promote the use of Facebook as 
a genuine social sharing platform, and be instrumental in the development of 
dense interaction clusters in the structure of the network. 

Figure~\ref{fig:fbb}(c) corresponds to the 15\% of users who 
are positioned at the periphery of a single alter that interacts with many
others.  Since the structure corresponds to an average or typical
ego-network structure for users in this role, it signifies a group of users who
are passive and seldom share information with others. When they do share,
it tends to be with those who the user has a mutual association with.  
Furthermore, these users tend to receive information from alters that share
prolifically. The phenomenon of over-active or extraordinarily well connected users on online social systems 
is well-studied~\cite{mislove07,kumar10,kwak10}, 
but it is interesting to discover that the users connected to them to also play
an important role in the online system. These users `absorb' the information of 
the over-active others, since they only forward such information to those
already connected to the over-active source. In fact, a modern
use of the Facebook platform is to ``absorb information" from friends 
and news organizations rather than to share social 
information, as reflected by this social role~\cite{baresch11,singer14,vu14}.

\subsection{Wikipedia}
\begin{table}
\centering
\begin{tabular}{c | c | c}
Role label & Structure & Proportion of users \\ \hline\hline
Interdisciplinary Contributor & Figure~\ref{fig:wikii}(a) & 89.7\% \\
Technical Editor & Figure~\ref{fig:wikii}(b) & 10.3\% \\
\end{tabular}
\caption{Wikipedia roles}
\label{tab:wikii}
\end{table}

\begin{figure}
\centering
\begin{tabular}{cc}
  \includegraphics[width=70mm]{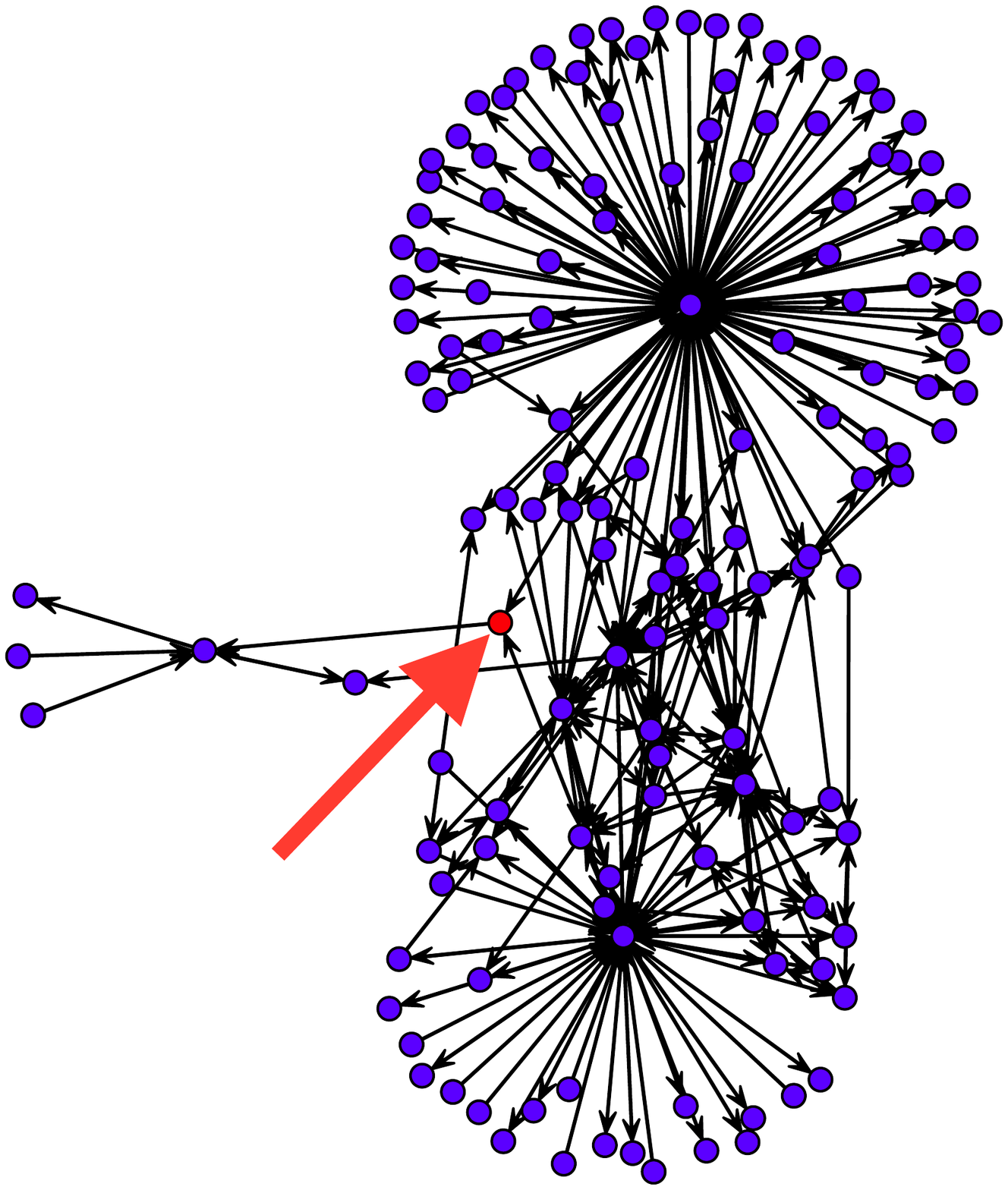} &
  \includegraphics[width=70mm]{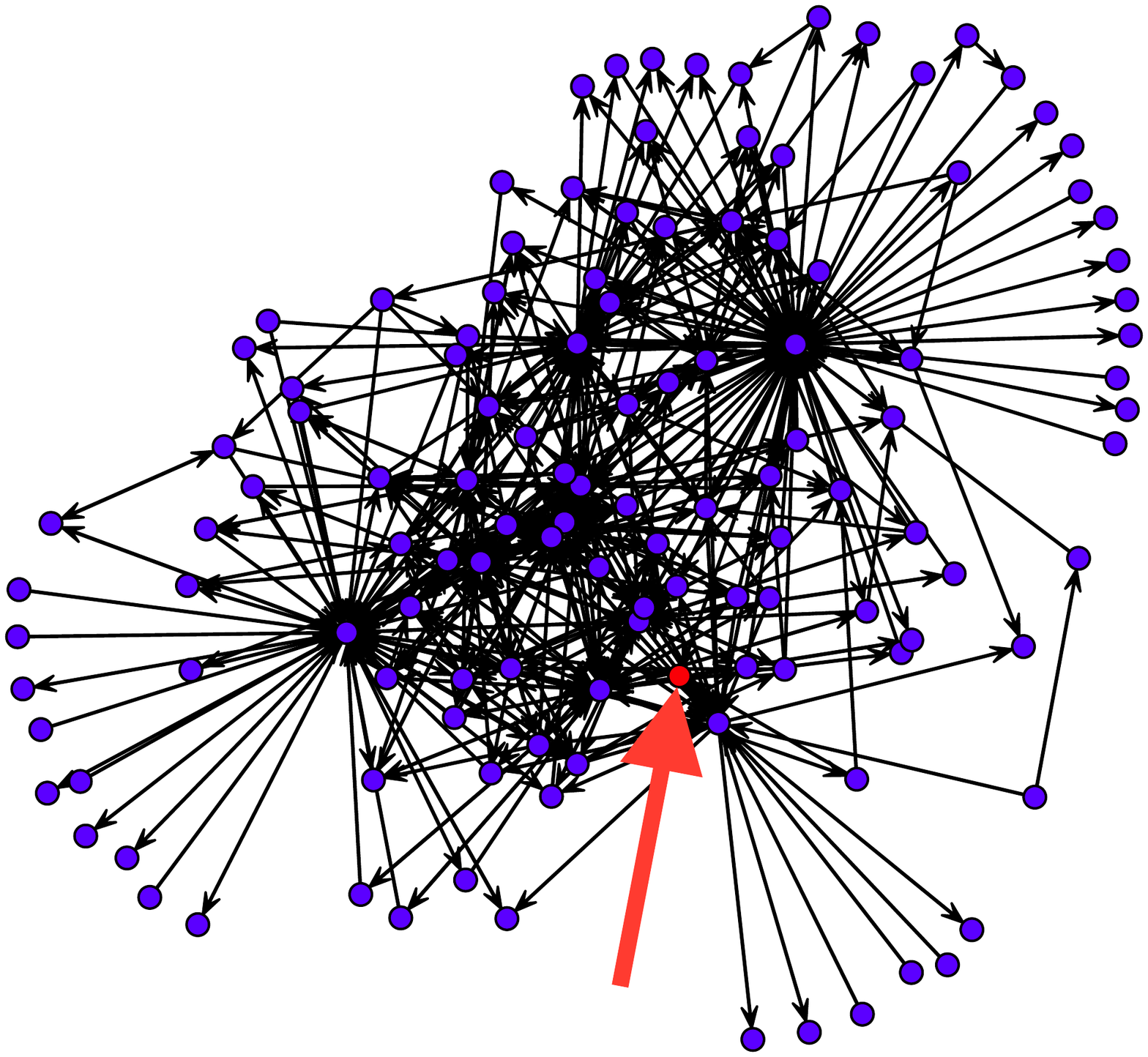} \\
  (a) & (b)
\end{tabular}
\caption{Central role structures: Wikipedia}
  \label{fig:wikii}
\end{figure}

The density of the central structures of the Wikipedia social roles 
shown in Figure~\ref{fig:wikii} is a result of the many different ways 
interactions are defined, which includes content editing, reverting a change,
or voting on a pending action by another user. The triad-based analysis
revealed two types of roles in Wikipeida. The first role is taken on 
by the majority of all users (89.7\%) and has the central structure
shown in Figure~\ref{fig:wikii}(a). The structure shows a user
whose work is being changed by active alters that make changes to articles from
many other authors as well. It is interesting that these the active alters
seldom edit content added by a common individual (e.g. have few mutual connections), even though they are prolific
editors. Such a pattern may emerge when these alters have different 
expertise and concentrate on editing contributions that fall within their 
specific domain. The existence of these `hubs' of editing activity is 
not a surprising finding, as 
past work has confirmed that most editors on Wikipedia do
exhibit domain-specific expertise and limit their edits articles in their
domain~\cite{welser11}. Users falling under this structure must therefore
be contributing to interdisciplinary articles, which most Wikipedia 
articles are classified as~\cite{medelyan08}. Such ``interdisciplinary
contributors" represent the vast majority of users (89.7\%) 
and is thus the primary role that adds information to Wikipedia.

The remaining 10.3\% of users fall under the role whose central structure is given in Figure~\ref{fig:wikii}(b). 
Two alters that take the
form of a hub (a domain-specific expert) can be seen, but the overlap between them
is larger and denser in comparison to Figure~\ref{fig:wikii}(a). The central user is positioned within this overlap. Users in this 
role therefore edit the contributions of many, 
and find their contributions edited by many others as well. A plausible 
explanation for finding a dense core between the positions of domain experts
is that they perform `general' edits 
that reflect the language, grammar, spelling, hyperlinking, and structure of 
articles. Changes made by these ``technical editors" may be further refined 
by a large number of other editors to further refine the technical discussion
or the presentation and language of an article. This explanation is 
compatible with past observations of users that concentrate on edits 
related to the language and format of an article~\cite{welser11}. \\
 
\noindent
In summary, the analysis demonstrates the use of conditional triad censuses to 
extract social roles from different types of online social 
systems. It naturally discovered social roles on Facebook corresponding to 
users who maintain connectivity across many disconnected social groups (``
social group managers"), 
who participate in well-connected groups 
(``exclusive group participants") that generate many social interactions,
and passive users that serve as an outlet for over-active others to share 
information with (``information absorbers'') and may use Facebook as a 
platform to receive news. The roles discovered on Wikipedia focuses on the 
nature of the user's contribution to the content of the 
online encyclopedia. A majority of users (``interdisciplinary contributors") 
are devoted to articles that attract the attention of editors focusing 
on different subsets of articles, which may correspond to the actions of a 
domain-specific expert. The attraction of many experts suggests that the 
article the central user focuses on is interdisciplinary in nature. A minority
of users (``technical editors") edit many articles at once, and have their
articles edited by many others as well. These users may thus be domain-specific
experts or could be users that apply general language and formatting changes
to many articles on the site. 

\begin{figure}
\centering
\begin{tabular}{ccc}
  \includegraphics[width=55mm]{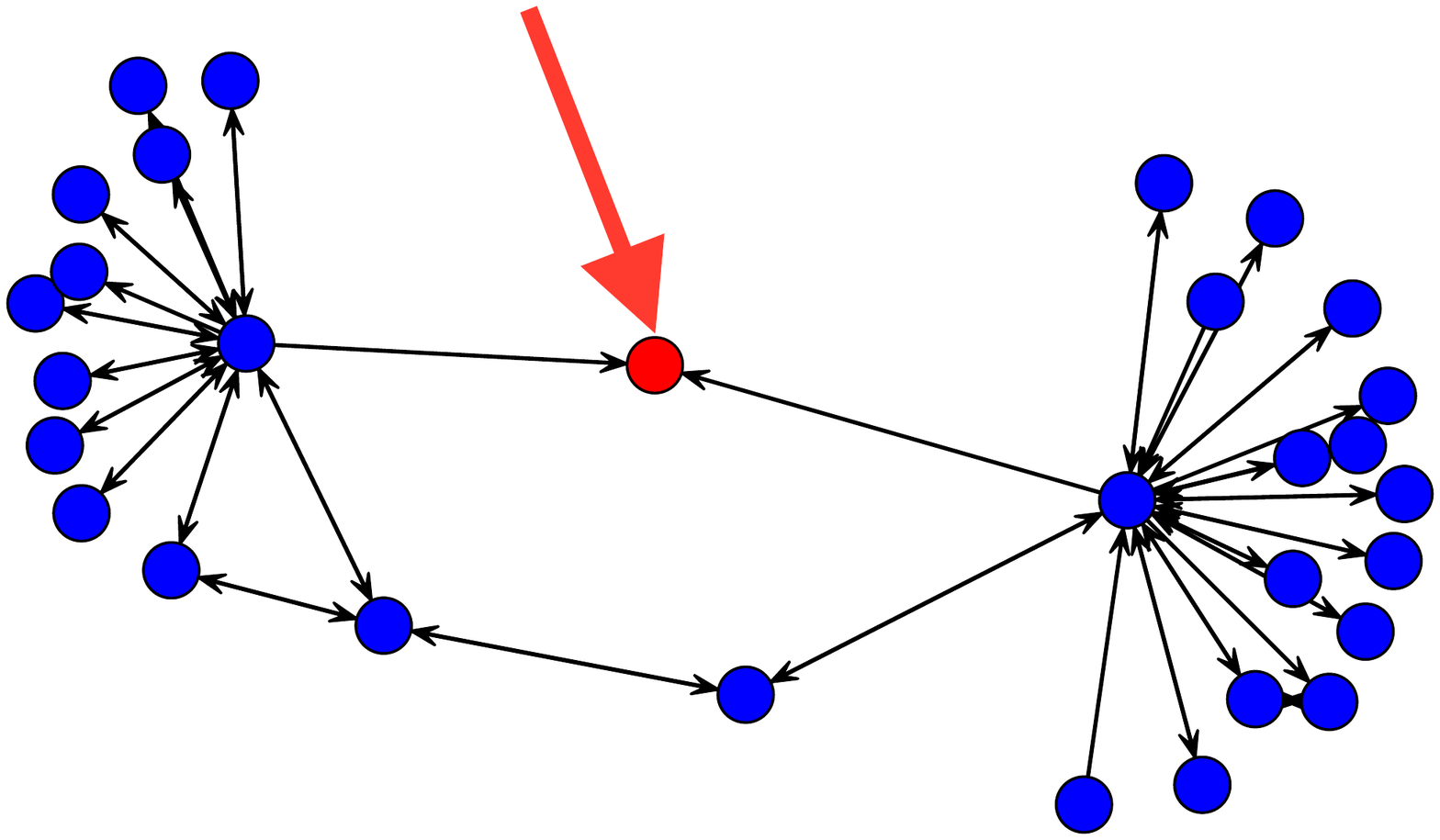} &
  \includegraphics[width=55mm]{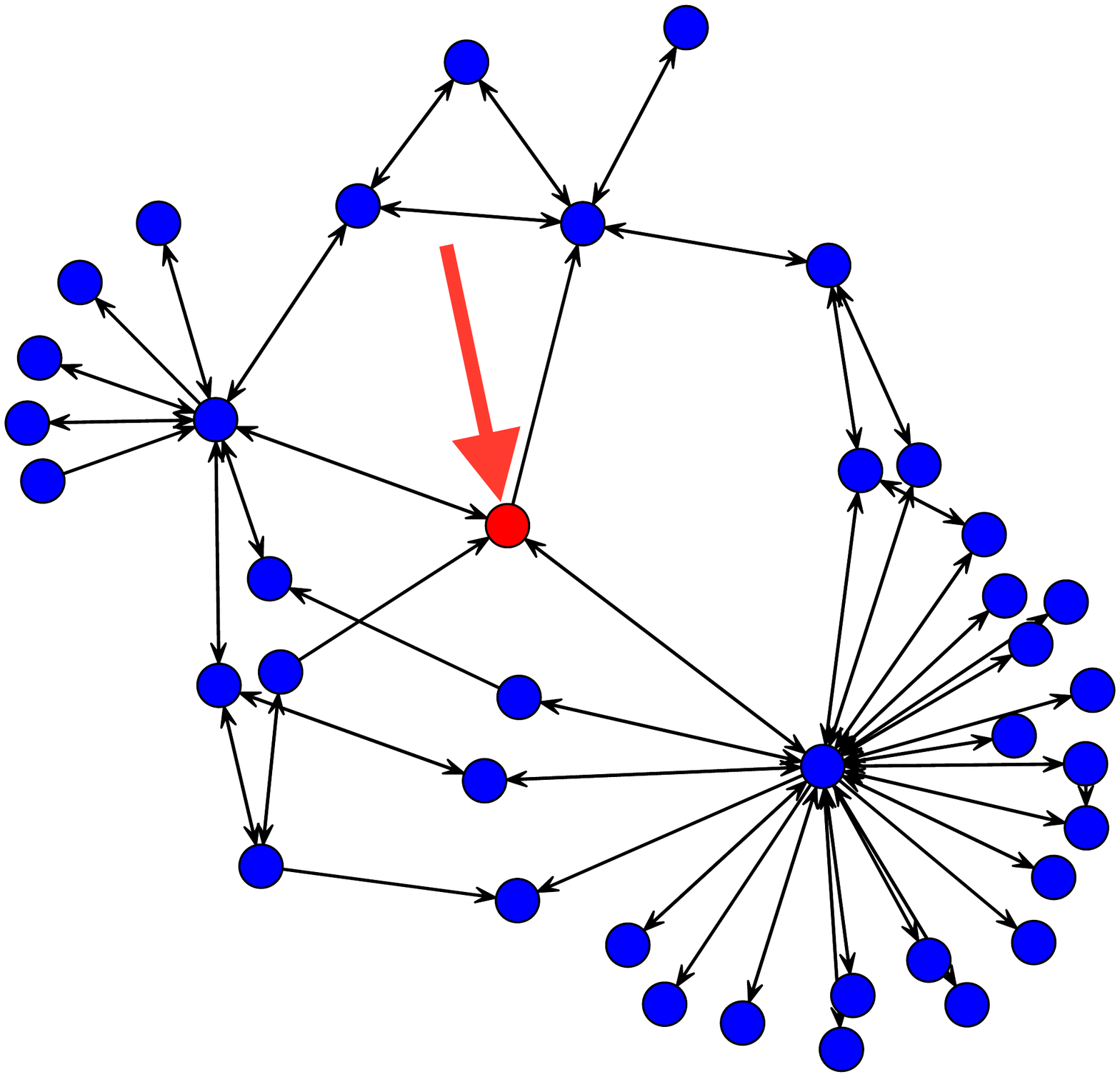} &
  \includegraphics[width=55mm]{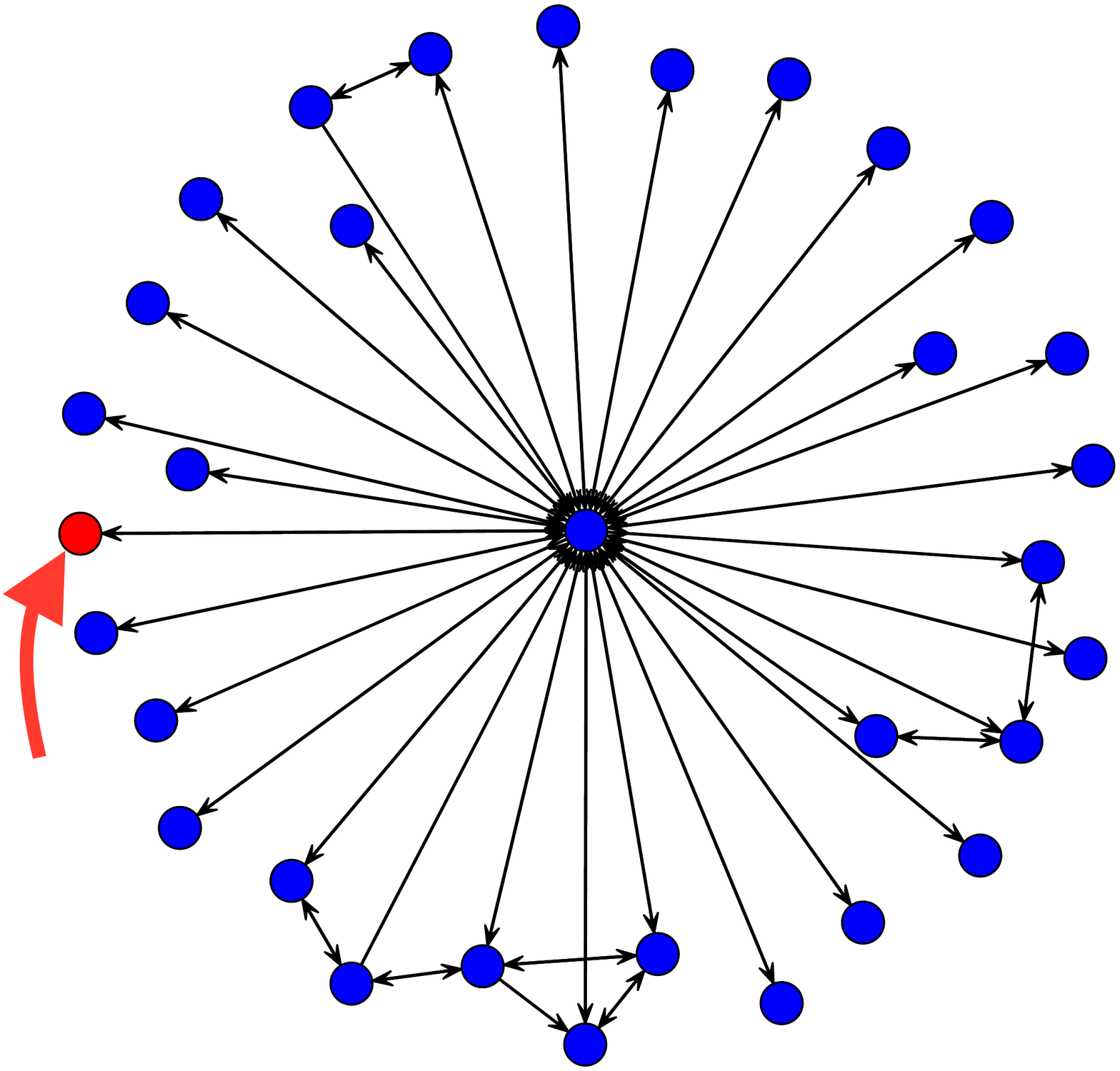}\\
    (a) & (b) & (c)  \\
\end{tabular}
\caption{Central role structures: UC Irvine online network}
  \label{fig:uci}
\end{figure}

\subsection{Applying social role analysis}
Triad-based social role analysis offers not only insights into
the nature of user behaviors on social systems, but also a practical tool for exploring
social theories. For example, 
consider a researcher wishing to study whether or not the reasons and ways users
interact with each other on Facebook is due to its inclusive,
public nature. This research question 
may be explored by comparing the social roles that emerge on Facebook with a different
online social network that is not inclusive and public, but exclusive and private. 
Differences in the number, shape 
and proportion of users falling into social roles across the two systems may 
give evidence of a relationship between the public or private nature of a social system and why
people participate in it. To illustrate this, a data set of interactions recorded from a 
private online social network for students at the University of California Irvine 
(UC Irvine) is considered~\cite{opsahl09}\footnote{It should be emphasized that a complete study of this research question requires a 
comprehensive analysis of user behaviors, and extensive comparisons 
between many different social network datasets. The illustration that follows is 
limited, and is only meant to demonstrate 
how social role analysis can be used as a useful research tool.}. The six-month long data set 
consists of 1,899 ties between users, with a directed tie from user $A$ to $B$
established when $A$ sends at least one message to $B$. 
Triad-based social role analysis on the UC Irvine network 
revealed the best clustering solution at $k=3$ roles ($\hat{C}_k = 0.713$).
Figure~\ref{fig:uci} visualizes the central structure of the resulting  role clusters, which
exhibit very similar features to the central structures of 
the social roles on Facebook. For example, Figures~\ref{fig:fbb}(a) and~\ref{fig:uci}(a)
both have a user situated between two  groups of others, Figures~\ref{fig:fbb}(b) and~\ref{fig:uci}(b)
find the user in the center of a well connected community, and Figures~\ref{fig:fbb}(c) and~\ref{fig:uci}(c)
shows the user sitting at the periphery of a highly active alter. 
An analyst may therefore consider the two networks to exhibit the 
same social roles, and hence, conclude users utilize the network for similar reasons
and in similar ways. Given the fact that Facebook and the UC Irvine social networks were created
to facilitate social communication and connection, it is not surprising to find similar roles and 
central structures emerging.

\begin{table}
\centering
\begin{tabular}{c  | c | c}
& \multicolumn{2}{c}{Proportion of Users} \\ 
 Role Label &  UC Irvine & Facebook \\ \hline\hline
Social Group Manager & 3.06\% & 56.6\% \\
Exclusive Group Participant &  92.9\% & 28.4\% \\
Information Absorber &  4.04\% & 15.0\% \\
\end{tabular}
\caption{Comparing the proportion of social roles on Facebook and UC Irvine networks}
\label{tab:ucir}
\end{table} 

Comparison of the shape and the proportion of users falling into the central role structures,
however,  reveal significant differences between the private UC Irvine and public Facebook online social
networks. For example, the central role structure of social group managers 
in the UC Irvine network finds the ego to be situated between 
a smaller number of groups compared to Facebook, and has an additional alter 
managing the same set of social groups.
These differences may arise because the separate groups an individual participates in within a 
private social network that is smaller in scope and encompasses fewer types of people 
may be less than a public social network that can include family, social, and work contacts. 
Furthermore, Table~\ref{tab:ucir} shows  the proportion of users falling into the social roles of the two
networks to be very different. The majority of users in the UC Irvine network are exclusive
group participants, that is, they are found to be embedded within a tight social group and do not need to manage 
a membership in many separate ones. In fact, only 3.06\% of UC Irvine users act as a social group manager, compared
to the 56.6\% of Facebook users that take on this role. 
This difference may be rooted in the fact that its users are
all students of UC Irvine, and hence, may exhibit homopholic tendencies
through common class, standing, housing, major, college, and club affiliations.
The many ways by which users could exhibit homophily on the UC Irvine network may also
explain why the social group manager central structure has an alter managing the same
set of groups as the ego; both could be managing groups of colleagues from the same 
class and club. The public nature of Facebook, however, may be reducing the level of homophily among
a user's connections. An analyst may point to these findings as key differences between public and private 
online social networks, and as a rationale to explore new hypotheses involving a comparison of 
homopholic tendencies within them. 

\section{Further Opportunities for Large Scale Social Role Analysis}
\label{sec:opp}
Based on the related
work discussed in Section~\ref{sec:rr} and on a reflection 
of the proposed triad-based method, this section summarizes 
additional challenges and opportunities that exist
in social role analysis for large scale social systems. Opportunities along 
two important directions are considered:~(i) finding meaningful features for 
role extraction; and~(ii) 
understanding the relationship between functional and social roles. 
	
\subsection{Linking representation with social theory}
As discussed in Section~\ref{sec:dda}, many data-driven analyses
select a large collection of structural, user, and
relationship attributes, and use them all to discover the social roles 
within a system. However, this may be a dangerous practice because the resulting
roles are defined to be according to the `similarity' of a complex mixture of 
many variables. Furthermore, many quantitative structural, user, and 
relationship features do not necessarily have a close correspondence 
to a sociological theory that is related to the concept of 
a user role. For example, structural features such as the
clustering coefficient or betweenness centrality 
of a user within her ego-network can quantify how clustered its structure is, 
but does not identify the telling patterns 
of the interactions within it. Analysis that use a large number 
of features thus lead to a separation of users into roles that must be
defined very broadly, or where ego-network structures within roles may be
discordant and have have few interpretable structural regularities. Some
methods using a large collection of features also apply 
post-processing steps to the resulting groups~\cite{chan10,zhu11}, 
which may further distort any interpretation of the extracted roles.  

This article takes a step toward the exclusive use of features that carry
a specific social interpretation. However, it may be the case that 
additional features associated with social theories 
may improve the fidelity of the method's results, or 
that a different unsupervised learning algorithm should be used. 
For example, Field {\em et al.} note the importance of preserving not
only interactions, but also affiliation information between users
in a social system to define their position~\cite{field06}. 
Such a concept may be operationalized in a richer dataset containing 
affiliation information, by incorporating similarity measures of the 
rows of a $g \times n$ binary incidence matrix whose $i^{th}$ row and 
$j^{th}$ column is 1 if user $j$ is affiliated with group $i$. 
Another related concept is the importance of social influence 
to the way it impacts a user's social role~\cite{friedkin97}. Fortunately,
there have been many measures proposed for quantifying influence
that may be integrated into the social role mining 
process~\cite{pallotti11,fujimoto12,chen09inf,li12,chen10,kempe03}.
It is these kinds of factors, instead of conveniently chosen structural
and user features, that should be considered when grouping users into 
social roles.


\subsection{Linking functional and social roles} 
In an offline setting, people can interact, converse, and exchange ideas
with each other in virtually innumerable ways. However, most large scale
social datasets come from online systems that only offer a limited number
of well-defined ways for people to interact with one another. It may be
intuitive to think that these modes of interactions, which reflect the 
{\em functional} ways users participate on the social system, are 
associated or have an effect on the social roles they go on to exhibit.
For example, the roles identified over Wikipedia in this article were more
closely related to the types of interactions allowed by the service
(as an expert editing content or a generalist editing language form).  
The functionality provided by Facebook may also have helped users fall
into specific social roles; for example, users only participating in a 
cooperative group of others may leverage the ability to choose
what friendships to accept on Facebook, so that the group they are
embedded in is cohesive. The idea that users can only interact with 
others in a limited number of ways is a unique property of online
social systems compared to offline ones. Thus, advanced features used
to discover social roles may also need to be associated with the different 
functionalities of an online social service, with values that reflect
what functions and how frequently they are used. Such features that 
are found to be `significant' across classes of users falling under the
same social role may signal an association between the functionality
of a social system and its social roles. 

\section{Concluding Remarks}
\label{sec:conc}
This article presented a methodology to discover social roles in large scale 
social systems. The data-driven approach, rooted in the representation of 
ego-networks as a conditional triad census and implemented with a simple
unsupervised learner was
applied to two different online social systems. Structural analysis of the
ego-networks falling closest to the center of clusters of users with similar
conditional triad censuses suggested the presence of users on Facebook that
exclusively manage disconnect social circles or participate in a highly
collaborative singular one. It also found how content posted on Wikipedia
may attract either the attention of a number of domain experts, or of 
multiple generalist editors. The data-driven approach was motivated by a 
comparative analysis of the existing equivalence based, implied, 
and data-driven role discovery methods that had been proposed. It concluded  
by suggesting the integration of social theories to derive features for role mining,
and approaches to link together the notion of what a user can do on a social
system with her social role on it. Future work should explore these opportunities, and may also consider 
unsupervised learners that allow users to fall into multiple role assignments. 
It is hoped that this important 
topic will continue to gain more attention in the computational social
network analysis and mining community.

\bibliographystyle{abbrv}
\bibliography{snam}
\end{document}